\documentclass{jfm}
\usepackage{graphicx}
\usepackage{epstopdf, epsfig}
\usepackage{color,makecell}

\shorttitle{Vortex-induced in-line vibration}
\shortauthor{E. Konstantinidis, D. Dorogi and L. Baranyi}

\title{Resonance in vortex-induced in-line vibration at low Reynolds numbers}

\author{Efstathios Konstantinidis\aff{1}
  \corresp{\email{ekonstantinidis@uowm.gr}},
  D\'aniel Dorogi\aff{2}  \and L\'aszl\'o Baranyi\aff{2}}

\affiliation{\aff{1}Department of Mechanical Engineering, University of Western Macedonia, Kozani 50132, Greece
\aff{2}Department of Fluid and Heat Engineering, University of Miskolc, 3515 Miskolc-Egyetemv\'aros, Hungary}

\begin{document}

\maketitle

\begin{abstract}
We present simulations of a circular cylinder undergoing vortex-induced  vibration  in-line with a free stream in conjunction with a theory for the fluid dynamics. Initially, it is shown that increasing the Reynolds number from 100 to 250 results in a 12-fold increase of the peak response amplitude at a fixed mass ratio of $m^*=5$. Subsequently, $m^*$ is varied from 2 up to 20 at a fixed Reynolds number of 180. The response amplitude as a function of the reduced velocity $U^*$ displays a single excitation region with peak amplitudes of approximately 1\% of the cylinder diameter, irrespectively of the $m^*$ value. The vibration is always excited by the alternating  shedding of single vortices. We develop a new model for the in-line fluid force, which comprises  an inviscid inertial force, a quasi-steady drag, and a  wake drag induced by vortex shedding. Our analysis shows that the wake drag appropriately captures a gradual shift in the timing of vortex shedding in its phase variation as a function of $U^*$ while the magnitude of the wake drag displays a resonant amplification within the excitation region.  We use the theory to illustrate why peak amplitudes, which occur when the vibration frequency is equal to the structural frequency in still fluid, do not depend on $m^*$, in agreement with our simulations as well as previous experiments at  Reynolds numbers higher than considered here. This new theory provides physical insight which could not be attained heretofore by employing  semi-empirical approaches in the literature.  
\end{abstract}

\begin{keywords}
\end{keywords}

\section{\label{sec:intro}Introduction}
The flow periodicity due to the vortex shedding from a bluff body exposed to a fluid stream can cause structural vibration if the body is flexible or it is elastically mounted, which is referred to as `vortex-induced vibration'.  In practical applications, compliant structures usually have degrees of freedom to move both along and across the incident flow. Much of the fundamental research on the problem has dealt with rigid circular cylinders as bluff bodies, constrained elastically so as to have a single degree of freedom to oscillate either in-line with a free stream (the streamwise direction) or transversely  (the cross-stream direction).   The circular cylinder spawns the characteristic that is not prone to galloping vibration and vortex-induced vibration occurs in its purest form. In an early review on vortex shedding and its applications, \cite{King1977}  noted that maximum in-line amplitudes  are approximately 0.2 diameters, peak-to-peak, or about one-tenth of the corresponding maximum cross-stream amplitudes. As a consequence, subsequent research has mostly concentrated on purely transverse vortex-induced vibration as attested in later reviews on the topic \citep[see, e.g.,][]{Bearman1984,Sarpkaya2004,Williamson2004,Gabbai2005,Bearman2011,Paidoussis}. Insight into the fundamentals of vortex-induced vibration can be gained by the complementary study of either in-line or transverse vortex-induced vibrations since both  share the same excitation mechanism. The present study deals with the in-line case.

\subsection{Characteristics of in-line free response  }

\begin{figure}\vspace{5mm}
\centerline{\includegraphics[width=0.63\textwidth]{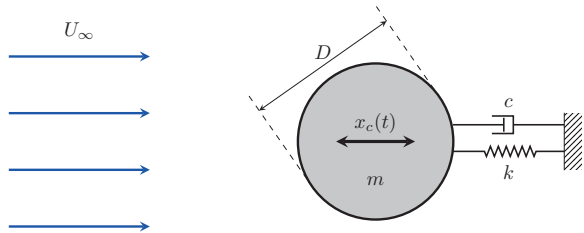}}
\caption{\label{fig:sketch} Schematic of the flow-structure configuration. }
\end{figure}

Figure \ref{fig:sketch} shows the flow-structure configuration considered in the present study. 
The elastically-mounted cylinder is modelled using the conventional mass-spring-damper  system. The  cylinder is constrained  so that it can  oscillate only  in-line with a uniform free stream. The motion of the cylinder is governed by Newton's second law, which can be expressed per unit span as 
\begin{equation}\label{eq:motion1}
	m\, \ddot{x}_c + c\, \dot{x}_c + k\, x_c = F_x(t),
\end{equation}
where $x_c$, $\dot{x}_c$, and $\ddot{x}_c$ respectively are the displacement, velocity and acceleration of the cylinder, $m$ is the mass of the cylinder, $c$ is the structural damping, $k$ is the spring stiffness, and $F_x(t)$ is the time-dependent sectional fluid force acting on the cylinder. 

The condition for the onset of vortex-induced in-line vibration can be broadly expressed as $U^*_a\approx 1/(2S)$, whereas the corresponding value for cross-stream vibration is $U^*_a\approx 1/S$, where $U_a^*= U_\infty/f_{n,a}D$ is the reduced velocity and $S=f_{v0}D/U_\infty$ is the Strouhal number; here $U_\infty$ is the velocity of the free stream, $D$ is the diameter of the cylinder, $f_{v0}$ is the frequency of vortex shedding for a stationary cylinder, and $f_{n,a}$ is the natural frequency of the structure in still fluid, i.e.\ including the `added mass'.  Traditionally, response data from experimental studies obtained as the flow velocity is varied over the attainable range of  experimental facilities are presented as a function of  the reduced velocity based on the structural frequency in still fluid. Other non-dimensional parameters governing the structural response are: the ratio of the cylinder mass to the fluid mass displaced by the cylinder, denoted as the mass ratio $m^*$, the ratio of the structural damping to the critical damping at which the mechanical system can exhibit oscillatory response to external forcing, denoted as the damping ratio $\zeta$, as well as the Reynolds number,  $Re$, which determines the flow regime.  The definitions of the mass ratio and the damping ratio are also not uniform in the literature but depend on whether the added fluid mass is taken into account. In this work, we have selected a set of non-dimensional parameters listed in table \ref{tab:parameters}. It should be noted that  we define  the reduced velocity using the natural frequency of the system  in vacuum, $f_n=(1/2\upi)\sqrt{k/m}$, in common with most previous numerical studies of vortex-induced vibration. 

\begin{table}
\begin{center}
\begin{tabular}{cccccc}
\thead{Normalized\\amplitude}   & \thead{Normalized\\frequency}   & 
 \thead{Reduced\\velocity} & \thead{Mass ratio} & 
\thead{Damping ratio} & \thead{Reynolds number} \\
$\displaystyle A^* = \frac{A}{D}$ & $\displaystyle f^* = \frac{fD}{U_\infty}$ & 
$\displaystyle U^* = \frac{U_\infty}{f_nD}$  & 
$\displaystyle m^* = \frac{m}{\frac{1}{4}\upi\rho D^2}$ & 
$\displaystyle \zeta = \frac{c}{2\sqrt{km}}$  &
$\displaystyle Re = \frac{\rho U_\infty D}{\mu}$  \vspace{3pt} \\
\end{tabular}
\caption{\label{tab:parameters} Definitions of  non-dimensional parameters employed in the present study.}
\end{center}
\end{table}

Typically, the response amplitude of cylinder vibration is magnified in distinct ranges of the reduced velocity,  which mimic the classical resonance of a single degree-of-freedom  oscillator to external harmonic forcing. These distinct regions of high-amplitude response have been given various names such as `instability regions' \citep{King1977},  `excitation regions' \citep{Naudascher1987}, or `response branches' \citep{Williamson2004}. 
It has been established as early as the 1970's  that there exist  two distinct excitation regions of free in-line vibration: the first one appears at $U_a^*\lesssim 2.5$  and has been associated with symmetrical shedding of  vortices simultaneously from both sides of the cylinder, whereas the second one appears at $U_a^*\gtrsim2.5$ and has been associated with alternating shedding of vortices from each side of the cylinder \citep{Wootton1972,King1977,Aguirre1977}. The value of $U_a^*\approx2.5$ corresponds to $1/(2S)$ assuming a Strouhal number of 0.20,  at which reduced velocity the frequency of vortex shedding from a stationary cylinder becomes equal to half the natural frequency of the structure in still fluid, i.e.\ $f_{v0}\approx \frac{1}{2}f_{n,a}$. A factor of 2 arises in the denominator from the fact that two vortices shed from alternate sides of the cylinder  each one induces a periodic oscillation of the fluid force in the streamwise direction.  It should be remembered that the Strouhal number is a function of the Reynolds number, defined as $Re=\rho U_\infty D/\mu$, where $\rho$ is the density and $\mu$ is the dynamic viscosity of the fluid. Therefore,  the above value of $U_a^*\approx2.5$  should generally be replaced by $U_a^*\approx 1/(2S)$ in dealing with response data at different Reynolds numbers.  

In experimental studies with elastically mounted rigid cylinders,  the structural frequency  depends on the oscillating mass and the stiffness of the supporting springs, which provide elastic restoring forces. 
In an early study, \cite{Aguirre1977} conducted more than a hundred tests in a water channel to investigate vortex-induced in-line vibration. He concluded that the structural mass and the stiffness  affected the cylinder response independently and in different ways: for a given value of $f/f_{v0}$, the density ratio (i.e.\ the mass ratio here) did not affect the  response amplitude when normalized with the cylinder diameter, nor did the stiffness affect the normalized frequency of vibration $f/f_{n,a}$,  where $f$ is the actual vibration frequency.  Beyond that early study, the effects of the structural mass and stiffness, which are embodied in the mass ratio and the reduced velocity values, have not  been systematically addressed in more recent studies. Yet, \cite{Okajima2004} found that the response amplitude of in-line oscillation decreases in both excitation regions with increasing the reduced mass-damping, or Scruton number - a non-dimensional parameter that is proportional to the product $m^*\zeta$ and is often employed to compile peak amplitude data as a function of a single parameter. The later findings  possibly illustrate the influence of structural damping alone since the mass ratio was constant in those tests. 

The existence of two distinct response branches of free in-line  vibration and corresponding modes of vortex shedding were confirmed in more recent experimental studies at Reynolds numbers in the range approximately from $10^3$ to $3.5\times10^4$ \citep{Okajima2004,Cagney2013a}. The drop in response amplitude in-between the two branches, i.e.\ at $U_a^*\approx2.5$, has been attributed to  the phasing of alternating vortex shedding, which provides a positive-damping,  or negative-excitation force with respect to the oscillation of the cylinder \citep{Konstantinidis2005,Konstantinidis2014}. More recently, a mixed mode of combined symmetric and alternating vortex shedding was also reported to exist  in-between the two branches \citep{Gurian2019}.

\subsection{Energy transfer and harmonic approximation}
For  self-excited vibrations to be possible, energy must be transferred from the fluid to the structural motion in an average cycle so as to sustain the oscillations, i.e.\ $E\geqslant0$ where $ E=\oint{F_x\mathrm{d}x_c}$; ${F_x}$ is the instantaneous fluid force driving the body motion and $x_c$ is the displacement of the body. A rationale is to determine the energy transfer from forced  vibrations of the body in order to predict whether free vibrations can occur for the corresponding case where the cylinder is elastically constrained. This is typically based on the  approximation that both the motion of the cylinder $x_c(t)$  and the driving fluid force per unit length $F_x(t)$ can be  expressed as single-harmonic functions of  time $t$, e.g.\
\begin{eqnarray}
		x_c(t) & = & X_0 + A\cos{(2\upi f t)},\label{eq:Xharmonic}\\
		F_x(t) & =  & F_{x0} + F_{x1}\cos{(2\upi ft+\phi_x)},\label{eq:Fharmonic}
\end{eqnarray}
where $X_0$ is the mean streamwise displacement of the cylinder, $A$ is the amplitude and $f$ is the frequency of body oscillation, $F_{x0}$ and $F_{x1}$ respectively are the magnitudes of the mean and unsteady in-line fluid forces, and $\phi_x$ is the phase lag between the displacement and the driving force at the oscillation frequency.  
Using the above harmonic approximations, it can be readily shown that 
\begin{equation}
	E = \upi A F_{x1}\sin\phi_x.
\end{equation}
Hence, free vibration is possible only if $\sin\phi_x\geqslant0$, or equivalently the phase lag be within the range $0^\circ\leqslant\phi_x<180^\circ$. 

Two independent studies employing forced harmonic in-line vibrations at fixed amplitudes of oscillation in the range from 0.1 to 0.28 diameters peak-to-peak, have shown that energy  is transferred from the fluid to the cylinder motion in two excitation regions separated by approximately $U^*_{r} \approx2.5$  for $Re$ values higher than $10^3$ \citep{Tanida1973,Nishihara2005}. Here, $U^*_{r} =U_\infty/fD$ is the reduced velocity based on the actual frequency of forced oscillation. The use of $U^*_{r}$  is also critical in correlating forced-vibration studies with the response  from free-vibration studies, in which case the vibration frequency is not necessarily equal to the structural frequency  \citep{Williamson2004,Konstantinidis2014}. Overall, predictions using forced harmonic vibration agree well with the excitation regions found in free vibration at relatively high Reynolds numbers, including the wake modes responsible for  free vibration \citep{Tanida1973,Nishihara2005}. On the contrary,  \citeauthor{Tanida1973} found that energy transfer  was always negative for all reduced velocities at $Re=80$.   
They stated that  results obtained for $Re=80$ were representative over the range $40\leqslant Re \leqslant150$,  which indicated that free vibration may not be possible in that range Reynolds numbers.   

More recently, detailed results from two-dimensional numerical simulations for a cylinder placed in an oscillating free stream and the equivalent case of a cylinder oscillating in-line with a steady free stream both showed that $E<0$ for all reduced velocities  at fixed Reynolds numbers of $Re=150$ \citep{Konstantinidis2017AOR} and $Re=100$ \citep{Kim2019}. The lowest amplitudes of forced oscillation in those studies were 0.1 and 0.05 diameters, respectively. The findings from both studies have also indicated that  free in-line  vibration may not be feasible for Reynolds numbers in the laminar  regime, which is consistent with the earlier experimental study of \cite{Tanida1973}. Nevertheless, it is plausible that energy transfer  may become positive at lower amplitudes of oscillation than  those employed in previous studies, which  has not received attention to date  since free in-line vibration has scarcely been studied at low Reynolds numbers; the authors are aware of  only few  numerical studies where in-line vibration of circular cylinder rotating at prescribed rates was investigated at $Re=100$ \citep{Bourguet2015,LoJacono2018}.  These studies showed that an elastically-mounted rotating cylinder can be excited into large-amplitude galloping-type vibrations as the reduced velocity increases. However, the response amplitudes were negligible  in the case of a non-rotating cylinder compared to the rotating cases and the former results were not discussed.   

Apart from addressing the question whether in-line vortex-induced vibration is possible at low Reynolds numbers, a more fundamental issue is to clarify what are the flow physics causing variations in the amplitude and frequency of response when self-excited vibration does occur, not only at low Reynolds numbers. This issue impacts our  understanding  of vortex-induced vibration as well as its modelling and prediction using semi-empirical codes in industrial applications. To address that issue, it is essential to formulate a theoretical framework with the aid of which results can be interpreted. In this study, we maintain that the in-line free vibration offers a convenient test case because it allows different  dynamical effects, which are associated with fluid inertia, fluid damping, and fluid excitation from the unsteady wake, to be segregated.

\subsection{Previous theoretical-empirical approaches}

A long-standing approach is to represent the in-line force per unit length $F_x(t)$ based on the equation proposed by \cite{Morison1950}. For a  cylinder of circular cross section oscillating in-line with a steady free stream, the equation can be written as 
\begin{equation}\label{eq:Morison1}
	F_x(t) = \frac{1}{2}\rho D C_{dh}\left|U_\infty - \dot{x}_c\right|\left(U_\infty - \dot{x}_c\right) -   \frac{1}{4}\upi\rho D^2C_{mh} \ddot{x}_c,
\end{equation}
 where $\rho$ is the density of the fluid. The coefficients $C_{dh}$ and  $C_{mh}$ are often referred to as drag and added mass (or inertia) coefficients, respectively, and their values are empirically determined from  measurements or simulations. Inherent to this approach is the harmonic approximation since the coefficients $C_{dh}$ and  $C_{mh}$ are often determined from tests where the cylinder is forced to vibrate harmonically. Even when the cylinder motion is self-excited, it is still necessary to characterize the vibration in terms of the least number of appropriate non-dimensional parameters for compiling fluid forcing data; this usually boils down to the use of two parameters, i.e.\ the normalized amplitude and normalized frequency of oscillation, which can fully characterize only single-harmonic oscillations. 

Another similar approach is to  decompose the fluctuating part of the in-line force into harmonic components in-phase with the displacement (or alternately acceleration) and in-phase with the velocity of the oscillating cylinder, in addition to a  steady term for the mean drag. 
By using harmonic approximations, the steady-state  response can be predicted as we present in  appendix \ref{app:harmonic}. This approach  is analogous to using Morison \etal's equation and linearising the drag term as  $\left|U_\infty - \dot{x}_c\right|\left(U_\infty - \dot{x}_c\right) \approx U_\infty^2 - 2U_\infty\dot{x}_c$. However, Morison \etal's equation comprises two force coefficients whereas the harmonic approximation comprises three force coefficients. The lack of an independent  term for the mean drag may explain, at least partially, why Morison \etal's equation  reconstructs the in-line force unsatisfactorily in the case of  vibrations in-line with a steady free stream.  However, it should be noted that  the addition of a third term for steady drag in Morison's equation did not considerably improve the empirical-fit results \citep[see][]{Konstantinidis2017AOR}.    

\citet{Sarpkaya2001} discussed some limitations of Morison \etal's equation to represent the in-line force on a cylinder placed perpendicular to zero-mean oscillatory flow.  Recently, \cite{Konstantinidis2017AOR} demonstrated the inability of the thus reconstructed force acting on a cylinder in non-zero-mean oscillatory flows to capture fluctuations due to vortex shedding in the drag-dominated regime, where the vortex shedding and the cylinder motion (the wave motion in that study)  are not synchronized. When the primary mode of sub-harmonic synchronization occurs, Morison's equation provides a fairly accurate fit to the in-line force but subtle differences still exist, which may have detrimental effects when the model equation is used for predicting the free response. A disadvantage of previous  approaches based on  Morison \etal's equation as well as on the harmonic representation, is that the values of the force coefficients  show some dependency on the best-fitting method, e.g.\ Fourier averaging \textit{vs.} least-squares method \citep[see][]{Konstantinidis2017AOR}.  Another disadvantage, more important, is that force coefficients are empirically determined and as a consequence it is difficult to decipher the flow physics from the variation of the force coefficients, which is our primary goal in this work.

\subsection{Force decomposition and added mass}
Despite the empirical use of Morison \etal's equation, its inventors stated that it originates from the summation of a quasi-steady drag force and the added-mass force resulting from `wave theory' \citep{Morison1950}. A comprehensive discussion of the theory can be found in \citet{Lighthill1986}. For a  body accelerating rectilinearly within a fluid medium, there is an ideal `potential' force acting on the body, which can be expressed as $F_{x,\mathrm{potential}}=-C_am_d\ddot{x}_c$, where $C_a$ is the added mass coefficient, $m_d$ is the mass of fluid displaced by the body, and $\ddot{x}_c$ is the acceleration of the body. Thus, the body behaves as if it has a total mass of $m+m_a$, where $m_a=C_am_d$ is the added mass of fluid. According to the theory of inviscid flow, in which case the velocity field can be defined by the flow potential, the added mass coefficient $C_a$ of any body  is assumed to depend exclusively on the shape of the body. For a circular cylinder, $C_a=1$.  Free-decay oscillation tests in quiescent fluid have shown that $C_a$ is quite close to the ideal value of unity.  However, the applicability of the ideal $C_a$ value in general flows, including cylinders oscillating normal to a free stream, has been criticized \citep{Sarpkaya1979,Sarpkaya2001,Sarpkaya2004}.  On the other hand, \citet{Khalak1996} argued that removing the ideal added-mass force from the total force will leave a viscous force that may still comprise a component in-phase with acceleration, i.e.\ the decomposition does not have to separate all of the acceleration-depended forces as done in empirical approaches. Ever since the separation of  `potential' (inviscid) and `vortex' (viscous) components has been widely employed  to shed light into the  vortex dynamics around   oscillating bodies transversely to a free stream \citep[see, e.g.,][]{Govardhan2000,Carberry2005,Morse2009b,Zhao2014,Zhao2018,Soti2018}.    

For body oscillations in-line with a free stream, one may also split the  streamwise force  as 
\begin{equation}
	F_x(t) = F_{x,\mathrm{potential}}(t) + F_{x,\mathrm{vortex}}(t),
\end{equation}
in order to explore the link between fluid forcing and vortex dynamics.
This was previously done  for the case of a fixed cylinder placed normal to a free stream with small-amplitude sinusoidal oscillations superimposed on a mean velocity \citep{Konstantinidis2011}. This case is kinematically equivalent to the forced vibration of the cylinder in-line with a steady free stream. In that study, large-eddy simulations corresponding to $Re=2150$ showed that alternating vortex shedding  provides positive energy transfer for $U^*_r>2.5$, in very good agreement with previous experimental studies discussed earlier. It was also observed that the streamwise vortex force diminished in magnitude while the instantaneous phase of the vortex force with respect to the imposed oscillation  drifted continuously near the middle of the wake resonance (synchronization) region. This was considered to be inconsistent with the flow physics in the following sense: within the synchronization region the vortex shedding and  the oscillation are strongly phase-locked  and  the corresponding wake fluctuations are resonantly intensified. Therefore, the magnitude of the vortex force would have been expected to increase and its instantaneous phase to remain fairly constant  in this region. The irregular phase dynamics observed in that study indicates that the  vortex force   remaining from  subtracting the ideal inertial force from the total force may not fully represent the effect of the unsteady vortex motions on the fluid forcing.

\subsection{New theory and outline of the present approach}
In this paper, we develop a new theoretical model for representing the streamwise force on a cylinder oscillating in-line with a free stream. The model stems from some recent observations. In particular, it was recently shown that Morison \etal's equation based on the sum of a quasi-steady viscous drag force and an inviscid inertial force represents the in-line force with comparable accuracy as does the equation with best-fitted coefficients over a wide range of parameters from the inertia to drag-dominated regimes \citep{Konstantinidis2017AOR}. However, neither method could capture fluctuations at the vortex shedding frequency in the drag-dominated regime, as noted earlier. Thus, the idea here is to introduce  an independent force term  $F_{dw}$, i.e.\ to express the total force as
\begin{equation}\label{eq:Morison2}
	F_x(t) = \frac{1}{2}\rho D C_{d}\left|U_\infty - \dot{x}_c\right|\left(U_\infty - \dot{x}_c\right) -   \frac{1}{4}\upi\rho D^2C_a \ddot{x}_c + F_{dw}(t),
\end{equation}
where the first term represents the quasi-steady drag, the second term represents the inviscid added-mass force, and the third term represents the unsteady force due to periodic vortex formation in the wake.  In this new approach, there are two viscous  contributions:  the quasi-steady drag, which is an `instantaneous' reaction force, and the wake drag, which  represents the `memory' effect in a time-dependent flow. These contributions may be thought of as originating from the vorticity  in the thin boundary and free shear layers, and from the vorticity  in the near-wake region, respectively. Both  are affected by the rate of diffusion of the vorticity, which is finite. However, at sufficiently high Reynolds numbers for which separation occurs,  the diffusion within the thin vortex layers occurs fast enough, almost `instantaneously'. Then, the force required to supply the rate of increase of the kinetic energy of the rotational motion in these regions may be taken to be  proportional to the square of the relative velocity $U_\infty-\dot{x}_c$, which gives rise to a quasi-steady drag \citep{Lighthill1986}. On the other hand, the diffusion of vorticity at the back of the cylinder is a very complex process involving its cross-annihilation as oppositely-signed vortices roll-up close together in the formation region \citep[see, e.g.,][]{Konstantinidis2016}; as a consequence the resulting fluid force acting on the body depends on the history of the vortex motions  in the near wake. 

The splitting of the viscous drag to quasi-steady and wake components is consistent with the contribution of vorticity in distinguishable flow regions around a cylinder to the fluid forces as shown in the  work of  \citet{Fiabane2011}. They revealed these separable contributions by displaying force-density distributions based on the volume-integral expression   proposed by \citet{Wu2007}. \citeauthor{Fiabane2011} were able to separate an  `external-flow' region containing the thin vortex structures in the attached and free shear layers, which contributed 90\% of the mean drag, and a `back-flow' region between these two vortex layers behind the cylinder, which contributed almost all the drag fluctuations.  Moreover, they found that the intensification of the vortex roll-up closer to the cylinder with increasing Reynolds number in the range $Re=50-400$ resulted an increase of the drag fluctuations imposed by the back-flow region. Their findings suggest that  -- when the cylinder is oscillating -- the `external flow' is at the origin of the quasi-steady drag  whereas the contribution from the `back flow' is captured by the wake drag. Interestingly, \citet{Wu2007} also considered the flow around a circular cylinder in a steady free stream as a test case in their study; they remarked that while a concentrated vortex after its feeding sheet is cut off  makes little direct contribution to the fluid force, it plays an indirect but major role through its induced effect on unsteadiness of the boundary-layer separation and the motion of separated shear layers, which implies that wake vortices influence the phasing of the fluid forces.

Assuming that a single periodic mode of vortex shedding occurs, the wake-induced force can be further modelled as a single-harmonic function of time, i.e.
\begin{equation}\label{eq:Fvharmonic}
	 		F_{dw}(t) = \frac{1}{2}\rho U_\infty^2DC_{dw}\cos{(2\upi f_{dw}t+\phi_{dw})},
\end{equation}
where the coefficient $C_{dw}$ represents the magnitude of the unsteady wake drag, and $\phi_{dw}$ the phase between the wake drag and the displacement of the cylinder. The frequency $f_{dw}$ excited by the unsteady wake depends on the vortex-shedding mode, e.g.\  $f_{dw}=2f_{vs}$ for the alternating mode, whereas $f_{dw}=f_{vs}$ for the symmetrical mode. 
Equation~(\ref{eq:Fvharmonic}) serves as a reduced-order model with the aid of which the fluid dynamics of vortex-induced vibration can be analysed more thoroughly than possible heretofore by employing previous semi-empirical approaches from the literature.

In this study, we conducted numerical simulations of the flow-structure interaction of a circular cylinder elastically constrained so as to oscillate only in-line with a free stream. The  main objectives are: (\textit{a}) use the numerical data from simulations to obtain the variations of the model parameters $C_{dw}$ and $\phi_{dw}$ as functions of the problem parameters, and  (\textit{b}) use the  expressions derived to calculate the model parameters in conjunction with the equation of cylinder motion to develop a theoretical framework for interpreting the phenomenology of vortex-induced in-line vibration. Simulations were restricted to the two-dimensional laminar regime at low Reynolds numbers to keep computer time within reason so as to examine the influence of the reduced velocity and the mass ratio over wide ranges and with a good resolution. The numerically produced sets of data for flow fields and induced fluid forces and their interaction with the resulting free motion of the cylinder allowed  us to address the issues raised in the foregoing paragraphs and hopefully make a contribution to the understanding of the complex flow physics.

\section{Methodology}

\subsection{Governing equations\label{sec:equations} }
The flow is assumed incompressible and two-dimensional while physical properties of the fluid are constant. The fluid motion is governed by the momentum (Navier--Stokes) and  continuity equations, which can be written in non-dimensional form   using the pressure-velocity formulation as
\begin{eqnarray}
	\frac{\partial u_x}{\partial t} + u_x\frac{\partial u_x}{\partial x} +  u_y\frac{\partial u_x}{\partial y} & = & - \frac{\partial p}{\partial x}  + \frac{1}{Re} \left(\frac{\partial^2 u_x}{\partial x^2} + \frac{\partial^2u_x}{\partial y^2} \right)  - \ddot{x}^*_{c},	\label{eq:Xmomentum}  \\
	\frac{\partial u_y}{\partial t}  + u_x\frac{\partial u_y}{\partial x} + u_y\frac{\partial u_y}{\partial y} &=& - \frac{\partial p}{\partial y}  +  
	\frac{1}{Re} \left(\frac{\partial^2 u_y}{\partial x^{2}} + \frac{\partial^2u_y}{\partial y^{2}} \right),\\
	 \frac{\partial u_x}{\partial x} + \frac{\partial u_y}{\partial y}&=& 0.\label{eq:continuity}
\end{eqnarray}
Coordinates are normalized with $D$, fluid velocities with $U_\infty$, time with $D/U_\infty$, and  pressure with $\rho U_\infty^2$.  The acceleration of the cylinder $\ddot{x}^*_c$ appears on the right-hand side of equation (\ref{eq:Xmomentum}) because the Navier--Stokes equations are applied in a non-inertial frame of reference that moves with the vibrating cylinder. Instead of explicitly enforcing the continuity equation, the  pressure field was computed by solving the following Poisson equation at each time step, 
\begin{equation}\label{eq:poisson}
\nabla^2p  = 2\left( \frac{\partial u_x}{\partial x}\frac{\partial u_y}{\partial y}   -   \frac{\partial u_x}{\partial y}\frac{\partial u_y}{\partial x} \right)  
	 -\frac{\partial \mathcal{D}}{\partial t},
\end{equation}
where $\mathcal{D}=\partial u_x/\partial x + \partial u_y/\partial y$ is the dilation. Although the dilation is zero by default in incompressible flows (Eq.~\ref{eq:continuity}), the term  $\partial\mathcal{D}/\partial t$ is kept in Eq. (\ref{eq:poisson}) to avoid the propagation of numerical inaccuracies \citep{Harlow1965}. 

On the cylinder surface, the no-slip boundary condition gives
\refstepcounter{equation}
$$
		u_x = 0, \qquad u_y = 0, \eqno{(\theequation{\mathit{a},\mathit{b}})}\label{eq35}
$$
and the following condition for the normal pressure gradient at the wall
\begin{equation}
	\frac{\partial p}{\partial n} = \frac{1}{Re}\nabla^2 u_n - \ddot{x}^*_{c,n},  
\end{equation}
where $n$ refers to the component normal to the cylinder surface pointing to the fluid side. 
At the far field, a potential flow field is assumed so that
\refstepcounter{equation}
$$
				u_x = u_{x,pot} - \dot{x}_c^*,  \qquad u_y = u_{y,pot},
				\eqno{(\theequation{\mathit{a},\mathit{b}})} \label{eq:BCfar}
$$
where $u_{x,pot}$ and $u_{y,pot}$ is the  velocity field from the known potential of irrotational flow. The corresponding condition for the far-field pressure  is
\begin{equation}
		\frac{\partial p}{\partial n} = 0. \label{eq:BCfar2}
\end{equation}
The initial field corresponds to the potential flow around a circular cylinder. The dimensionless form of the equations illustrates that the fluid motion depends solely on the Reynolds number given the boundary and   initial conditions. 

The displacement of a cylinder elastically constrained  so that it can  oscillate only  in-line with a uniform free stream is governed by Newton's law of motion, which can be written in non-dimensional form as 
\begin{equation}\label{eq:motion}
	\ddot{x}_c^* + \frac{4\upi\zeta}{U^*}\dot{x}_c^* + \left(\frac{2\upi}{U^*}\right)^2 x_c^* =	\frac{2C_x(t)}{\upi m^*},
\end{equation}
where $x_c^*$, $\dot{x}_c^*$, and $\ddot{x}_c^*$ respectively are the non-dimensional displacement, velocity, and acceleration of the cylinder, normalized using $D$ and $U_\infty$ as length and velocity scales; $C_x(t)$ is the sectional fluid force on the cylinder normalized by $0.5\rho U^2_\infty D$. Here, the fluid forcing is provided by the ambient flow through balancing the normal and shear stresses on the cylinder surface. The cylinder is initially at rest. The dimensionless form of equation (\ref{eq:motion}) illustrates that the cylinder motion depends on the reduced velocity $U^*$, the mass ratio $m^*$, and the damping ratio $\zeta$. The full set of independent dimensionless parameters of the problem comprises $Re,\,U^*,\,m^*$, and $\zeta$.  

Because of the two-dimensional approach, simulations were limited up to $Re = 250$. 
Our main simulations are at  $Re = 180$, a value which is slightly lower than the threshold for which mode-A spanwise instability occurs in the wake of a stationary cylinder, i.e.\ $Re_c \approx190$ \citep[see][]{Williamson1996,Barkley1996}, and may provide some indication of the corresponding threshold of three-dimensional transition for oscillating cylinders. Yet, the in-line oscillation of the cylinder causes wake synchronization, which has been shown to suppress the  mode-A spanwise instability into two-dimensional laminar flow with strong K\'arm\'an vortices at $Re=220$ \citep{Kim2009}.  Therefore, the flow may well be expected to remain strictly two dimensional for  our main simulations at $Re=180$. 
For simulations at the highest value of $Re=250$,  it is plausible that some three-dimensional instability might exist. Three dimensionality usually appears first in the form of weak coherent structures of streamwise vorticity with specific wavelength riding on the primary spanwise vorticity that remains in-phase along the length of freely-vibrating cylinders \citep[see][]{LoJacono2018,Bourguet2020}.  Under such flow conditions, the spanwise vorticity component is much higher than the other two components by one order and the direct effect of the three-dimensionality of the vortical structures on the force is weak as noted by \citet{Wu2007}.   Thus, the plausible existence of three-dimensional vortex structures in the cylinder wake may be reasonably expected to not directly influence the magnitude and phase of the streamwise and transverse fluid forces, which are primarily determined by the two-dimensional wake instability, nor influence the cylinder response at the highest Reynolds number of 250 at which we conducted simulations to check the trends of results in the laminar regime. 

\subsection{Numerical code}
An in-house code based on the finite difference method was used to solve the equations of fluid motion \citep{Baranyi2008}. The flow domain is enclosed between two concentric circles: the inner circle is the boundary fitted to the cylinder surface while the outer circle represents the far field boundary. The polar physical domain is mapped into a rectangular computational domain using linear mapping functions. The computational mesh of the `physical domain' is fine in the vicinity of the cylinder and coarse in the far field while the corresponding mesh of the transformed domain is equidistant. Space derivatives are approximated using a fourth order finite-difference scheme except for the convective terms for which a third-order modified upwind difference scheme is employed. The pressure Poisson equation is solved using the successive over-relaxation (SOR) method and the continuity equation is implicitly satisfied at each time step. The Navier-Stokes equations are integrated explicitly using the first-order Euler method and the fourth-order Runge--Kutta scheme is employed to integrate the equation of cylinder motion in time. At each time step the fluid forces acting on the cylinder are calculated by integrating the pressure and shear stresses around the cylinder surface, which are obtained from the flow solver. The streamwise force is supplied to the right-hand side of Eq.~(\ref{eq:motion}) which is integrated to advance the cylinder motion. At the next time step, the cylinder acceleration is updated and the equations of fluid motion are integrated to complete the fluid-solid coupling. For all simulations reported here,   the cylinder is initially at rest and the initial field around the cylinder satisfies the potential flow.

\subsection{Domain size, grid resolution and time-step dependence studies}
In this section, we present  results from preliminary simulations to check the  dependence of main output parameters  on a) the size of the computational domain in terms of the radius ratio $R_2/R_1$, b) the grid resolution $\xi_{max}\times\eta_{max}$, and c) the dimensionless time step $\Delta t$. Here $\xi_{max}$ and $\eta_{max}$ are the number of grid points in peripheral and radial directions, respectively.  During these computations,  the Reynolds number, the reduced velocity, the mass ratio, and the structural damping ratio  were fixed at  $Re=180$, $U^*=2.55$, $m^*=10$, and $\zeta=0$, respectively. The main output parameters of interest are the amplitude $A^*$ and frequency $f^*$ of cylinder response  (normalized with $D$ and $U_\infty$) as well as the standard deviations of the in-line and transverse fluid forces (normalized with $0.5\rho U_\infty^2D$), which are respectively denoted $C'_x$ and  $C'_y$.

\begin{table}
\begin{center}
\def~{\hphantom{0}}
\begin{tabular}{lccllll}
domain & $R_2/R_1$ & $\xi_{max}\times\eta_{max}$ & \mbox{$A^*$} & \mbox{$f^*$} & \mbox{$C'_x$}  & \mbox{$C'_y$} \\
small &120 & $360\times274$  & 0.01082 & 0.3730 & 0.06974 & 0.4226 \\
medium &160 & $360\times291$  & 0.01079 & 0.3726 & 0.07051 & 0.4231 \\
large & 200 & $360\times304$  & 0.01077 & 0.3725 & 0.07094 & 0.4235 \\
\end{tabular}
\caption{\label{tab:Indep_Area} Results of the domain dependence studies at $(U^*, m^*, Re)=(2.55, 10, 180)$}
\end{center}
\end{table} 

\begin{table}
\begin{center}
\def~{\hphantom{0}}
\begin{tabular}{lcllll}
grid  & $\xi_{max}\times\eta_{max}$ & \mbox{$A^*$} & \mbox{$f^*$} & \mbox{$C'_x$}  & \mbox{$C'_y$} \\
coarse & 300$\times$242 & 0.01074 & 0.3725 & 0.07087 & 0.4235 \\
medium & 360$\times$291 & 0.01079 & 0.3726 & 0.07051 & 0.4231 \\
fine & 420$\times$339 & 0.01082 & 0.3727 & 0.07033 & 0.4229 \\
\end{tabular}
\caption{\label{tab:Indep_Grid} Results of the grid dependence studies at $(U^*, m^*, Re)=(2.55, 10, 180)$}
\end{center}
\end{table} 

\begin{table} 
\begin{center}
\def~{\hphantom{0}}
\begin{tabular}{cllll}
$\Delta t$ & \mbox{$A^*$} & \mbox{$f^*$} & \mbox{$C'_x$}  & \mbox{$C'_y$} \\
0.0004 & 0.01082 & 0.3727 & 0.07060 & 0.4233 \\
0.0002 & 0.01079 & 0.3726 & 0.07051 & 0.4231 \\
0.0001 & 0.01078 & 0.3726 & 0.07049 & 0.4231 \\
\end{tabular}
\caption{\label{tab:Indep_Time} Results of dimensionless time step dependence studies  at $(U^*, m^*, Re)=(2.55, 10, 180)$}
\end{center}
\end{table} 

First,  three different values of the radius ratio $R_2/R_1$ of the inner and outer circles defining the computational domain were tested and the corresponding results are shown in table~\ref{tab:Indep_Area}.  The number of circumferential nodes $\eta_{max}$ of the physical domain was adjusted in each case  in order to keep the grid equidistant in the transformed domain. For these computations,  the dimensionless time step was fixed at $\Delta t=10^{-4}U^*\cong0.0002$. Table~\ref{tab:Indep_Area} shows that the most sensitive quantity on the domain size is  $C'_x$ displaying a relative difference of 1.7\%  between the small and large domains, whereas the corresponding differences for $A^*$, $f^*$ and $C'_y$ are below 0.7\%. The relative differences for all quantities of interest become less than 0.6\% between the medium and large domains. Thus, the medium-sized domain with a radius ratio of $R_2/R_1=160$ was chosen for the rest of the computations.

Next,  we tested three  grids with different resolutions $\xi_{max}\times\eta_{max}$  where the number of peripheral and radial grid points was increased so that the grid remains equidistant in the transformed plane. For these computations, the radius ratio and dimensionless time step values were fixed at $R_2/R_1=160$ and $\Delta t=10^{-4}U^*\cong0.0002$, respectively. The results of these tests are shown in table~\ref{tab:Indep_Grid}. Again, $C'_x$ is the most sensitive quantity displaying a relative difference of 0.7\% between coarse and fine grids.  All quantities of interest display relative differences less than 0.3\% between  medium and fine grids. Thus, the medium-resolution grid was chosen for further computations.

Finally, we tested the dependence on the dimensionless time step, $\Delta t$ for three different values  corresponding  to $\Delta t\cong2\times10^{-4}U^*, 10^{-4}U^*$ and $5\times10^{-5}U^*$. For these computations, the radius ratio and grid resolution values were fixed at $R_2/R_1=160$ and $360\times292$, respectively. The results are shown in table~\ref{tab:Indep_Time}. In this tests, the quantity that is most sensitive to the time step is  $A^*$, which shows a relative difference of 0.4\% between the largest  and smallest time steps whereas the corresponding relative differences for $f^*$, $C'_x$ and $C'_y$ are all below 0.15\%. The relative differences of  all quantities of interest  are below 0.1\% between the intermediate and the smallest time steps. Thus,  a dimensionless time step of $\Delta t=10^{-4}U^*$ was chosen for the main computations.

\subsection{Code validation}

\begin{figure}
\centerline{\includegraphics[width=0.95\textwidth]{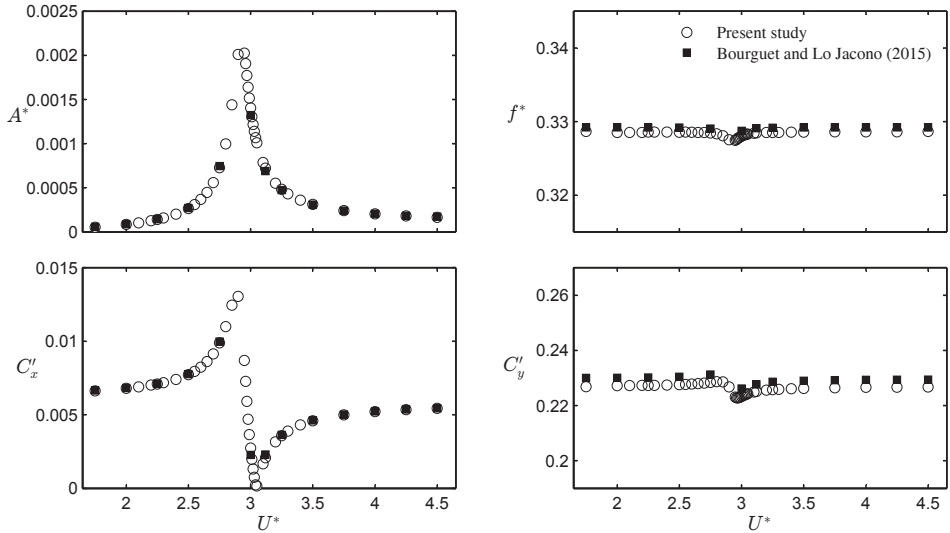}}
\caption{Comparison of results obtained in the present study (open circles) against the study of \cite{Bourguet2015} (filled squares) pertinent to purely in-line free vibration in terms of the variation of normalized amplitude $A^*$ and normalized  frequency $f^*$ of cylinder response and the standard deviations of the normalized forces in-line and transverse to the free stream,  $C'_x$ and $C'_y$ respectively, as functions of the reduced velocity $U^*$ for $(Re,\,m^*,\,\zeta)=(100,\,4/\upi,\,0)$.} \label{fig:comparison} 
\end{figure}

The computational code used in the present study was previously employed in several studies of flows about stationary and oscillating  cylinders and results have been extensively compared against data from the literature \citep[see][]{Baranyi2008,Dorogi2018,Dorogi2019}. 
For instance, \cite{Baranyi2008}  found good agreement with the study of  \cite{Al-Mdallal2007} in terms of the time history of the lift coefficient and Lissajous patterns of lift  \textit{vs.} cylinder displacement at comparable situations for the case of a cylinder forced to oscillate in the streamwise direction. In addition, the extended code handling flow-structure interaction has been validated for the case of a cylinder undergoing vortex-induced vibration with two degrees of freedom with equal natural frequencies in the streamwise and transverse directions against  results from \cite{Prasanth2008} for $m^*=10$, $\zeta=0$, and Reynolds numbers in the range from 60 to 240 \citep[for details see][]{Dorogi2018}. Furthermore, \cite{Dorogi2019} showed that results obtained with the present code compare well with published results in \cite{Navrose2017} for purely transverse free vibration, as well as in \cite{Prasanth2011} and  \cite{Bao2012} for free vibration with two degrees of freedom with equal or unequal, respectively, natural frequencies in the streamwise  and transverse directions, for similar conditions in each case. 

In addition to the validation tests presented in previous studies, we compare  in figure \ref{fig:comparison} results obtained with the present code against the study of \cite{Bourguet2015}  for purely in-line free vibration. Here, we employed a finer step in the reduced velocity to  resolve  the maximum in $A^*$ as well as the minimum in $C'_x$.
There is excellent agreement of results for $A^*$ and  $C'_x$  but there are some minor deviations for $f^*$ and  $C'_y$ of 0.2\% and 1.2\%, respectively, which might be attributable to  different numerical methods employed in those studies (finite difference \textit{vs.} spectral element). Overall, previous and present validation tests show that the numerical code employed in the present study provides accurate solutions.

\section{Results and discussion}
 
\subsection{Effect of Reynolds number on in-line response at a fixed mass ratio}

\begin{figure}
 \centerline{\includegraphics[width=0.68\textwidth]{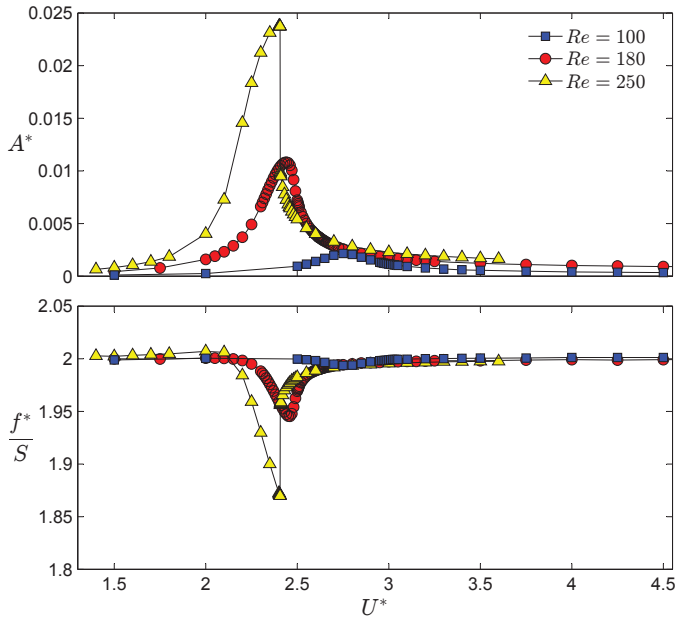}}
\caption{The in-line response amplitude $A^*$ and frequency $f^*$ with reduced velocity $U^*$ at different Reynolds numbers (see legend); $(m^*,\,\zeta)=(5,\,0)$. The $f^*$ values are divided by the corresponding $S$ values to take into account the effect of Reynolds number on the Strouhal number.} \label{fig:Re_response} 
\end{figure}

To start with, we consider  effect of  the Reynolds number on the in-line response of a cylinder with a mass ratio of $m^*=5$. The structural damping was set to zero so as to allow for the highest possible amplitude response to take place. 
Figure~\ref{fig:Re_response} (top plot) shows that there exists a single excitation region in which the response amplitude $A^*$ displays a marked peak for all Reynolds numbers considered. The $U^*$ value at which peak amplitudes occur decreases with $Re$, which can be attributable to the corresponding increase of the Strouhal number  since peak amplitudes occur at approximately $U^*_a\approx 1/(2S)$. The peak amplitude over the entire $U^*$ range, denoted as $A^*_\mathrm{max}$, increases from 0.002 at $Re=100$ to 0.024 at $Re=250$, i.e.\ an increase in $Re$ by a factor of 2.5 results in an increase  in $A^*_\mathrm{max}$ by a factor of 12. This is a remarkable increase of the order of magnitude, which contrasts the constancy of peak amplitudes of purely transverse free vibration in the corresponding range of Reynolds numbers \citep[a compilation of $A^*_\mathrm{max}$ data as a function of $Re$ from several studies can be found in][]{Govardhan2006}. 

For all simulations conducted in this study, the vortex shedding remained synchronized at half the frequency of the cylinder oscillation. As shown in the bottom plot in figure~\ref{fig:Re_response} the normalized response frequency  $f^*$ is approximately twice the corresponding Strouhal number at each Reynolds number. However, $f^*$ displays a trough within the excitation region that becomes more pronounced as the Reynolds  number increases; the trough can be hardly discerned for $Re=100$. The variations of $A^*$ and $f^*$ appear to be strongly correlated. The characteristics of free response remain similar for all Reynolds numbers considered here. However, a sudden drop in $A^*$ appears just after the peak amplitude for $Re=250$, a feature which is not present for lower Reynolds numbers. This might indicate the onset of branching behaviour similar to that observed in vortex-induced vibration purely transverse to the free stream \citep[see][]{Leontini2006}.   

\begin{figure}
 \centerline{\includegraphics[width=0.96\textwidth]{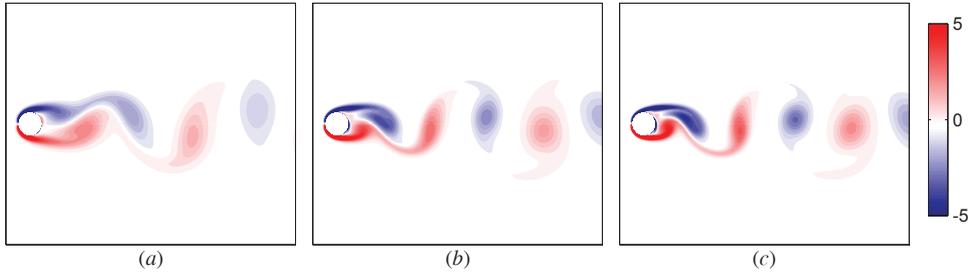}}
\caption{\label{fig:Re_wake} Snapshots of the distribution of vorticity around a cylinder undergoing free in-line vibration for $(m^*,\,\zeta)=(5,0)$; (\emph{a}) $(Re,\,U^*)=100,\,2.75)$, (\emph{b}) $(Re,\,U^*)=(180,\,2.44)$, (\emph{c}) $(Re,\,U^*)=(250,\,2.40)$. $U^*$ values correspond to peak response amplitudes  in figure \ref{fig:Re_response}. Each snapshot corresponds to a  random phase of the cylinder oscillation.  Contour levels of normalized vorticity at $\pm0.1,\pm0.5,\pm0.9,\ldots$.}
\end{figure}

In comparison to previous experimental studies, which typically correspond to Reynolds numbers above $10^3$, we did not observe another excitation region associated with symmetrical vortex shedding. In contrast, we observed only the alternating mode of vortex shedding in the present simulations corresponding to  the laminar wake regime. Figure~\ref{fig:Re_wake} shows  vorticity distributions  in the wake at $U^*$ values  corresponding to peak amplitudes for different Reynolds numbers. In all cases, the vorticity  distributions display the familiar von  K\'arm\'an vortex street similar to the wake of a stationary cylinder. 
As the Reynolds number is increased, the contours of individual vortices become more concentrated and peak vorticity values within them increase. This might be partly attributable to the increase in the amplitude of cylinder oscillation with Reynolds number, which accrues the generation of vorticity on the cylinder surface \citep{Konstantinidis2016}. In addition, the streamwise spacing between the centres of subsequent vortices decreases due to the increase of the normalized frequency of cylinder oscillation $f^*$ with Reynolds number. The absence of the other excitation region associated with the symmetrical vortex shedding may be attributable to the fact that,  as has been shown in several previous studies where the cylinder is forced to oscillate in the streamwise direction at correspondingly low Reynolds numbers, the onset of this mode occurs at  relatively high amplitudes above 0.1 diameters \citep{Al-Mdallal2007,Marzouk2009,Kim2019}. Since streamwise amplitudes of free vibration are much lower than that threshold, it is not surprising that the mode of symmetrical shedding and the corresponding excitation region were not observed in the present study.

\subsection{Effect of mass ratio on in-line response at a fixed Reynolds number}
Next, we concentrate on the effect of mass ratio on the in-line response at a fixed Reynolds number of $Re=180$, at which the flow is expected to remain laminar and strictly two dimensional. The structural damping was set to zero $(\zeta=0)$ to allow the highest possible amplitudes. 

\subsubsection{Cylinder response }
\begin{figure}
\centerline{\includegraphics[width=0.72\textwidth]{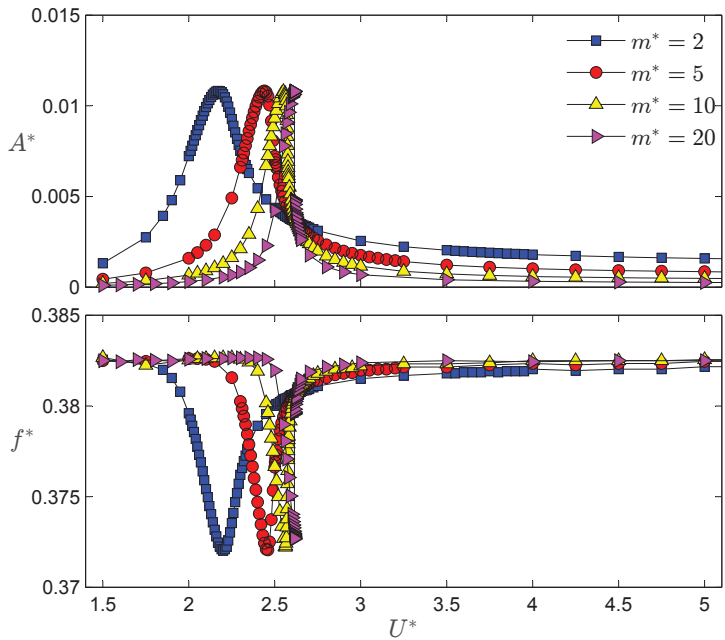}}
\caption{\label{fig:response} The variation of normalized amplitude $A^*$ and normalized frequency $f^*$ of cylinder response as  functions of the reduced velocity $U^*$ for $Re=180$ and different mass ratios $m^*$ (see the symbol legend for $m^*$ values). }
\end{figure}

Figure~\ref{fig:response} shows the variations of $A^*$ and $f^*$ with $U^*$ for four  $m^*$ values. It can be seen that $A^*$ displays a single excitation region with peak amplitudes of approximately 1\% of the cylinder diameter, irrespectively of $m^*$. At high reduced velocities, i.e.\  $U^*>4$, the response amplitude  gradually drops off  down to a level that depends on the mass ratio, with $A^*$ becoming lower  as $m^*$ increases. 
The response frequency $f^*$ initially decreases within the excitation region  reaching a minimum  value of approximately 0.372 for all mass ratios.  As  $U^*$ is increased beyond the point of minimum, $f^*$ increases asymptotically  towards the value corresponding to twice the Strouhal number for a stationary cylinder, i.e.\ $f^*\approx2S=0.384$.  It is interesting to note that  $f^*<2S$ over the entire $U^*$ range for all $m^*$, i.e.\ the cylinder always oscillates at a  frequency slightly lower than twice the frequency of vortex shedding from a stationary cylinder. This is consistent with forced harmonic vibration studies, which  show that vortex-induced vibration due to alternating vortex shedding occurs for $f<2f_{v0}$ \citep{Nishihara2005,Konstantinidis2011}.

The variations of $A^*$ and $f^*$ shown in figure~\ref{fig:response} appear to be strongly correlated, i.e.\ $A^*$ increases as $f^*$ decreases and \textit{vice versa}. However, it should be noted that peak amplitudes occur at marginally lower $U^*$ values than those that correspond to the minimum in $f^*$.  The  $U^*$ values at which peak amplitudes occur depend  on $m^*$ quite substantially. In fact, peak amplitudes occur approximately at the point where the frequency of cylinder oscillation approaches the natural frequency of the system in still fluid, i.e.\ $f\approx f_{n,a}$. This condition can also be expressed in dimensionless form as:
\begin{equation}\label{eq:Upeak}
	\mathrm{Normalized~frequency~at~point~of~peak~amplitude:}\qquad f^*\approx\frac{1}{U^*}\sqrt{\frac{m^*}{m^*+C_a}}.    
\end{equation}
The above condition can be verified in table~\ref{tab:Apeak}, which summarizes the response characteristics at peak amplitudes for different mass ratios. Effectively, we see that  the reduced velocity  at which the peak amplitude occurs primarily depends on the mass ratio but the peak response amplitude  does not depend on the mass ratio. 

\begin{table}
\begin{center}
\small\addtolength{\tabcolsep}{10pt}
\begin{tabular}{cllll}
$m^*$ & \mbox{$U^*$} & \mbox{$A_\mathrm{max}^*$} & \mbox{$f^*$}  &  $\frac{1}{U^*}\sqrt{\frac{m^*}{m^*+1}}$ \\
2 & 2.17 & 0.01081 & 0.3725  & 0.3763\\
5 & 2.44 & 0.01082 & 0.3724  & 0.3741\\
10&2.55  & 0.01078 & 0.3726  & 0.3739\\
20& 2.614 & 0.01081& 0.3727 & 0.3733\\
\end{tabular}
\caption{\label{tab:Apeak} Response characteristics at peak amplitude for different mass ratios ($Re=180)$. }
\end{center}
\end{table}

The drop in normalized frequency $f^*$ within the excitation region in fact illustrates the tendency of the oscillation to `lock-in' at the natural frequency of the structure in still fluid, i.e.\ $f\approx f_{n,a}$, over a  range of reduced velocities. It is important to distinguish this `lock-in' tendency, which only occurs  in free vibration, from `vortex lock-in' (a.k.a.\ `vortex lock-on'), which regards the synchronization of the vortex shedding and the cylinder oscillation and may occur in both forced and free vibration \citep{Konstantinidis2014}.  In forced vibration, there is no  natural frequency of the structure so the lock-in relationship $f=f_{n,a}$ is meaningless. It should also be noted that for all simulations reported in  figure~\ref{fig:response} (i.e.\ for all $U^*$ and $m^*$ values), the vortex shedding was synchronized with the cylinder oscillation so that $f=2f_{vs}$, where $f_{vs}$ is the frequency of  vortex shedding from the freely vibrating cylinder. This is tantamount to the  sub-harmonic vortex lock-on  in the context of forced oscillations where alternating vortex shedding synchronizes at half the frequency of cylinder oscillation, i.e.\ $f_{vs}=\frac{1}{2}f$ \cite[see, e.g.,][]{Kim2019}. However, the conventional lock-in $f\approx f_{n,a}$ occurs over a narrow range of reduced velocities in the region of peak-amplitude response.

\subsubsection{Magnitude and phase of fluid forces}

\begin{figure}
\centerline{\includegraphics[width=0.74\textwidth]{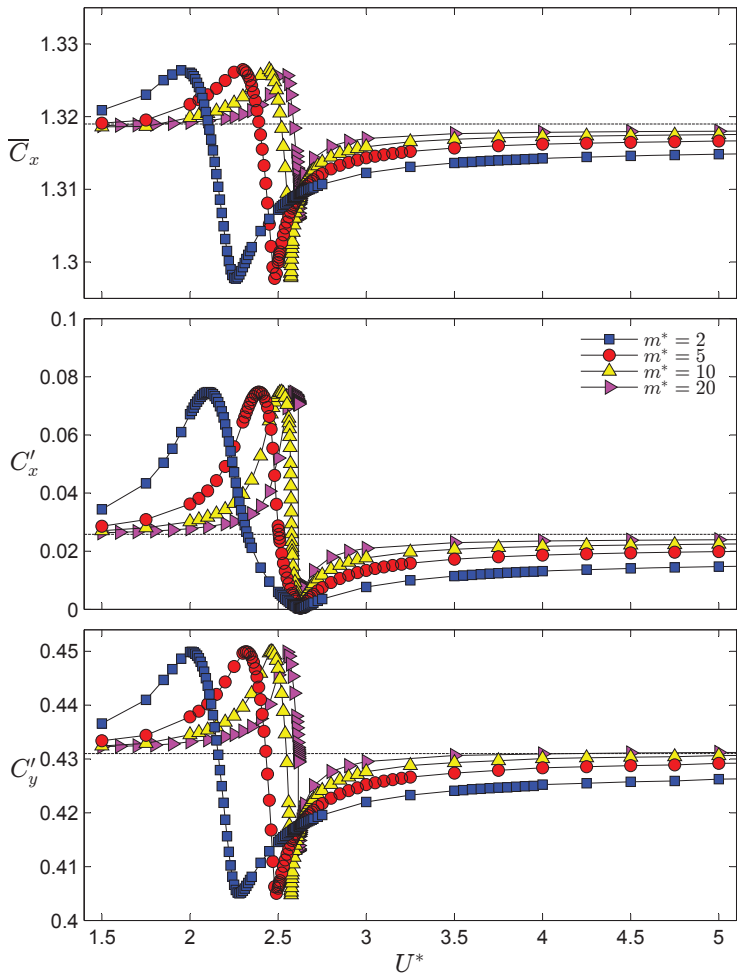}}
\caption{The variations of  the mean drag coefficient $\overline{C}_x$ and the standard deviations  of the normalized fluid forces  in-line with and transversely to the free stream, respectively $C'_x$ and $C'_y$, as  functions of the reduced velocity $U^*$  for  different mass ratios, $m^*$ at $Re=180$ (see the  legend for $m^*$ values). Dashed lines indicate constant values that correspond to the stationary cylinder.}\label{fig:forces}
\end{figure}

In figure \ref{fig:forces} we present the variations of the mean drag coefficient $\overline{C}_x$ and the standard deviations of the unsteady forces in-line with and transversely to  the free stream,  $C'_x$  and $C'_y$ respectively, as functions of  $U^*$. The dashed lines indicate constant values corresponding to a stationary cylinder at $Re=180$ \cite[taken from][]{Qu2013}. 
All three quantities exhibit similar variations with $U^*$  for all mass ratios, i.e.\ initially they  increase above the corresponding fixed-cylinder values, subsequently they  decrease steeply, and finally they gradually increase thereafter as  $U^*$ is increased. 
The maximum $\overline{C}_x$ is 0.5\% greater and the minimum $\overline{C}_x$ is 1.6\% smaller than the mean drag coefficient for the stationary cylinder,  independently of the $m^*$ value.  However, the maxima and minima in $\overline{C}_x$ occur at different $U^*$ values depending on the $m^*$ value.
The middle plot in figure~\ref{fig:forces} shows that  $C'_x$ reaches a peak value at approximately the point of peak amplitude response,  which is  nearly three times the value corresponding to a stationary cylinder;  a substantial  increase despite the low  amplitude of peak oscillation of only 1\% of the cylinder diameter.  
Following the peak there is  a steep decrease within a narrow $U^*$ range at the end of which $C'_x$ tends to zero. Interestingly, for all mass ratios  this  occurs at $U^*=2.625$, a point at which the response frequency coincides with the natural frequency of the structure in vacuum, i.e.\ $f=f_n$ (or in non-dimensional parameters $f^*=1/{U^*}$). Hereafter, this special operating point will be referred to as the `coincidence point' and its ramifications will be discussed in more detail in Sect.~\ref{sec:superharmonic}.    Beyond the coincidence point, $C'_x$  gradually reaches to a plateau at a value that is proportional to $m^*$. On the other hand, the peak and trough $C'_y$  values are merely 6\% higher and 5\% lower, respectively, than the value corresponding to a stationary cylinder (see the bottom plot in figure~\ref{fig:forces}). For all $m^*$, a maximum value of $C'_y=0.45$ exactly is attained at the point of peak response amplitude. In addition, at the coincidence point, i.e.\ $U^*=2.625$, $C'_y$ attains exactly the same value of 0.418  for all $m^*$, which is just slightly below the value corresponding  to a stationary cylinder. Since there is no body acceleration in the transverse direction, changes of $C'_y$ can  be related to changes in the vortex dynamics around the oscillating cylinder \citep{Leontini2013}. Then, the small variation in the $C'_y$ magnitude illustrates that  the process of vortex formation and shedding does not substantially change  over the entire range of reduced velocities, even though the cylinder can be oscillating with different but generally small amplitudes. 

\begin{figure}
\centerline{\includegraphics[width=0.74\textwidth]{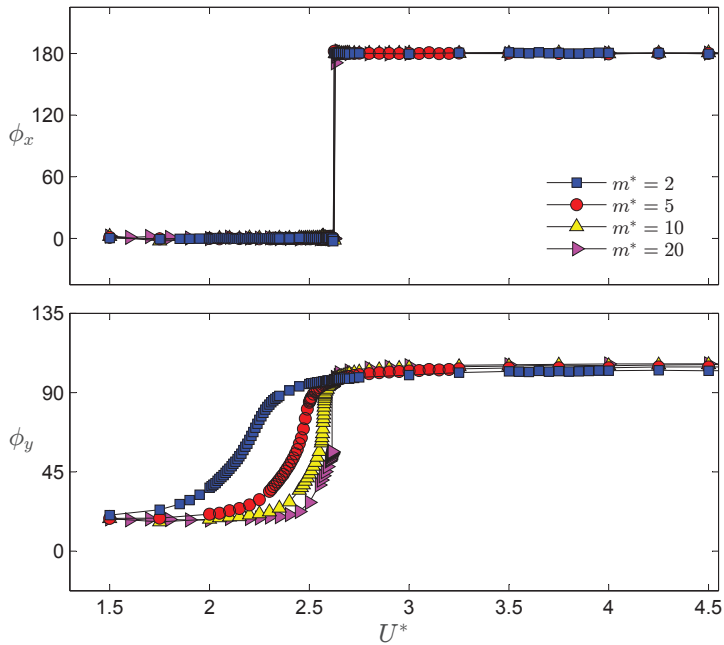}}
\caption{The variation of the phase angles $\phi_x$ and $\phi_y$ between in-line and transverse forces, respectively, and the cylinder displacement as  functions of the reduced velocity $U^*$ at $Re=180$ and different mass ratios $m^*$ (see the symbol legend for $m^*$ values).}\label{fig:phases}
\end{figure}

The phase angle $\phi_x$ between the driving force $F_x(t)$ and cylinder displacement $x_c(t)$, as defined by  harmonic approximations in equations  (\ref{eq:Xharmonic}) and (\ref{eq:Fharmonic}), is often considered to be a useful parameter in studies of vortex-induced vibration. In addition to $\phi_x$,  we also compute here the phase angle $\phi_y$ between the unsteady force acting in the transverse direction $F_y(t)$ and the displacement, assuming that $F_y(t)$ can also be approximated as a single-harmonic function of time. The calculation method of the phase angle is described in appendix \ref{app:phases}. It should be noted that the harmonic approximations work very well at low Reynolds numbers considered here except for the time history of $F_x(t)$ near the coincidence point as  will be discussed in detail further below. 
Figure \ref{fig:phases} shows the variations of the phase angles $\phi_x$ and $\phi_y$ as functions of $U^*$. 

In figure \ref{fig:phases}, it can be seen  that $\phi_x$ jumps suddenly from $0^\circ$ to $180^\circ$ across the coincidence point at $U^*=2.625$ for all $m^*$. This behaviour can be predicted from the steady-state harmonic solution, which is given in appendix \ref{app:harmonic}, as follows.  Equation   (\ref{eq:Hsine}) requires that the condition $\sin\phi_x=0$ be satisfied when  the structural damping is null $(\zeta=0)$. This constrains $\phi_x$ to be either $0^\circ$ or $180^\circ$.  This taken in tandem with the requirement per equation (\ref{eq:Hcosine})  that $\cos\phi_x$ must change from positive to negative as $U^*$ increases through $f^*U^*=1$  translates to the jump in $\phi_x$ from $0^\circ$ to  $180^\circ$ appearing at the coincidence point. We would like to stress that the equation of cylinder motion constraints $\phi_x$, which makes it impossible to infer any changes in the flow  from the variation of $\phi_x$ as  the reduced velocity is varied.  

In contrast to $\phi_x$, there is no analogous constraint on $\phi_y$, which  instead of a jump  displays a smooth variation with $U^*$, as can be seen in figure \ref{fig:phases}. For all mass ratios, $\phi_y$ increases from  an initial value  of approximately $20^\circ$ at low reduced velocities to a terminal value of approximately $105^\circ$ at high reduced velocities, with precise values being slightly depended on $m^*$. The $U^*$ range over which $\phi_y$   changes rapidly is consistent with the  range of relatively high-amplitude response, which broadens as $m^*$ decreases (see figure \ref{fig:response}). The variation of $\phi_y$ clearly suggests a gradual change in the vortex dynamics as $U^*$ is varied over the prescribed range. Based on evidence from previous studies \citep[see][]{Konstantinidis2005,Konstantinidis2011}, our main hypothesis is that the variation of  the phase angle $\phi_y$ can be linked to a gradual shift in the timing of vortex shedding with respect to the cylinder oscillation as $U^*$ is increased. 
This hypothesis will be verified in the following subsection by inspection of vorticity distributions at different phases of the cylinder oscillation.

\subsubsection{Vorticity distributions}

\begin{figure}
\centerline{\includegraphics[width=0.99\textwidth]{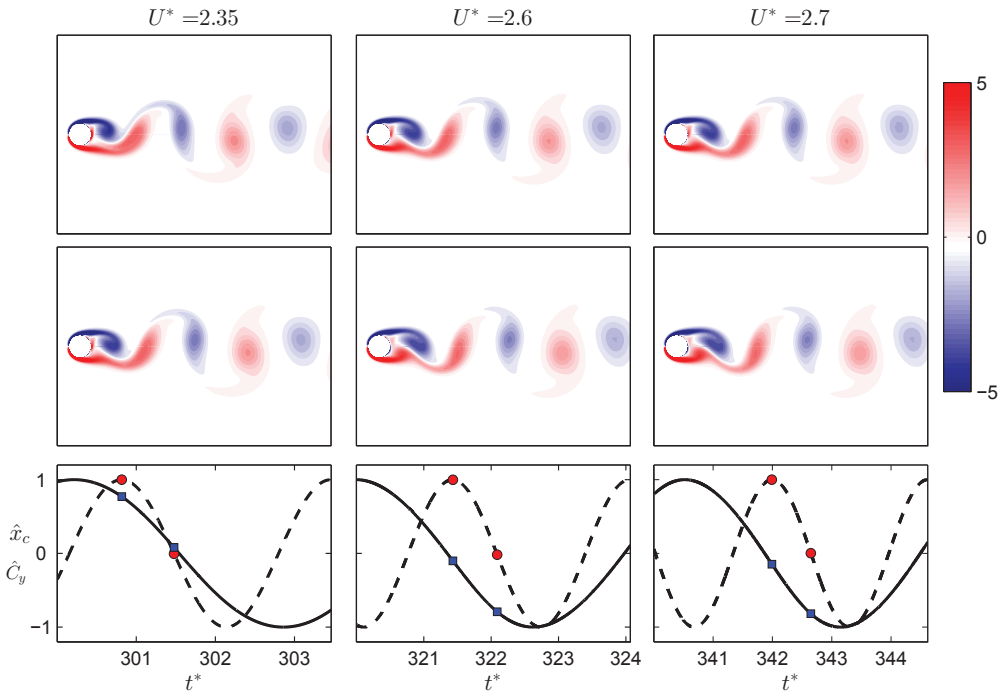}}
\caption{The top two rows show instantaneous distributions of the vorticity at maximum and subsequent zero displacement  during a cycle of cylinder oscillation  for three reduced velocities in each column;  $(Re,\,m^*)=(180,\,5)$.
The bottom row shows time traces of the normalized displacement $\hat{x}_c(t)$ (dashed lines) and the normalized transverse force $\hat{C}_y(t)$ (solid lines) where circle and square symbols, respectively, mark the instants for which vorticity distributions are shown above. Contour levels of normalized vorticity: $\pm0.1,\,\pm0.5,\,\pm0.9, \ldots$} 
\label{fig:vortex_phase} 
\end{figure}

We have selected three $U^*$ values, which are listed in table~\ref{tab:phase}, for presenting vorticity distributions in the wake. In purpose, the normalized frequency of cylinder oscillation $f^*$  is nearly the same in all three cases  so that the vorticity patterns  are easier to describe since the streamwise spacing of the shed vortices scales with $f^*$ \citep{Griffin1978}.  However, the frequency ratio $f/f_n=U^*f^*$ and the phase angle $\phi_y$ both increase with  $U^*$, at different degrees,  as also shown in  table~\ref{tab:phase}. The two higher $U^*$ values are just before and after the `coincidence point', $U^*f^*=1$. 
For each $U^*$ value, figure~\ref{fig:vortex_phase} shows instantaneous vorticity distributions at two phases corresponding to the maximum and the subsequent zero displacement of the cylinder as indicated in the time traces (see bottom plots in figure~\ref{fig:vortex_phase}). The significant change of the phase angle $\phi_y$ as a function of $U^*$ can be readily inferred from the time traces of the displacement and the transverse force, which have been appropriately normalized with their maximum amplitudes for better visualisation of their waveforms. 
As $U^*$ is increased, the position of individual vortices at corresponding instants appears to have shifted slightly downstream. The shift in the streamwise position of individual vortices    fits very well with the change in $\phi_y$  as $U^*$ is varied; when $U^*$ changes from 2.35 to 2.6, $\phi_y$ more than doubles and a large shift is observed; when $U^*$ changes from 2.6 to 2.7, $\phi_y$ changes by few degrees and the shift is hardly perceptible. 

We made a quantitative estimation of the relative phase shift for pairs of $U^*$ values from the corresponding spatial shift in the position of corresponding vortex centres at the instant of maximum displacement, with vortex centres  extracted by locating the points of peak vorticity in each individual vortex.  
When $U^*$ changes from 2.35 to 2.6 the above method yields a relative phase shift of $54^\circ$, which is very close to $\Delta\phi_y=55.4^\circ$ computed directly from the phase shift of the transverse force (see table \ref{tab:phase}). Similar inferences can be made by looking into the stage of  formation of vortices just behind the cylinder. For instance, a stripe of negative vorticity (in blue colour) connecting the second  vortex to the third vortex taken from right to left, i.e.\ as time progresses, can be observed at maximum displacement for $U^*=2.35$. However, the  corresponding vortices are no longer connected by a stripe of vorticity for  $U^*=2.6$ and 2.7, which illustrates that  vortex shedding is at a progressed stage in these cases.   
It should be noted that although this visual feature, which we could make clear by selecting a minimum level of the normalized vorticity of 0.1,  depends on this minimum level, it does provide a   quantitative comparison of the process of vortex formation at different reduced velocities since the same contour levels were employed for producing all vorticity distributions; in a similar manner the space occupied by individual vortices provides a measure of their strength, which allows quantitative comparisons when the same contour level is employed. It is also interesting to observe that nothing special changes in the vortex patterns  between   $U^*=$ 2.6 and 2.7 but a marginal phase shift, although $\phi_x$ jumps by  $180^\circ$ on crossing over the coincidence point.  
The above observations firmly support that the variation of  $\phi_y$ as a function of $U^*$, in contrast to the variation of $\phi_x$, is directly linked to changes in the timing of vortex shedding from the cylinder. 

\begin{table}
\begin{center}
\small\addtolength{\tabcolsep}{22pt}
\begin{tabular}[1.8\textwidth]{rlll}
\mbox{$U^*$} & 2.35 & 2.60 & 2.70 \\
\mbox{$f^*$}  &  0.378 & 0.380 & 0.381\\
\mbox{$U^*f^*$} & 0.888 & 0.989 & 1.030\\
\mbox{$\phi_y$} (degrees) & 40.6 & 96.0 & 99.2\\
\end{tabular}
\caption{\label{tab:phase} Important characteristics at three reduced velocities $(Re,\,m^*)=(180,\,5)$. }
\end{center}
\end{table}

\subsubsection{Variable added mass}
Changes in the frequency of cylinder response are often  correlated with variations in the inertial force due to the added mass . This follows from the relationship
\begin{equation}\label{eq:CGW}
	\frac{f}{f_{n,a}}=\sqrt{\frac{m^*+C_a}{m^*+C_{EA}}}.
\end{equation}
Here, $C_a$ is the ideal added mass coefficient from potential flow theory and $C_{EA}$ is a variable added mass coefficient \citep{Aguirre1977}. The latter coefficient has become known as the `effective added mass' in the context of transverse free vibration \citep{Khalak1996,Williamson2004}. Equation (\ref{eq:CGW}) is equivalent to (\ref{eq:Hcosine}) of the harmonic approximation solution where the component of the force in-phase with displacement has been substituted by
\begin{equation}
	C_{x1}\cos\phi_x = 2\upi^3f^{*2}A^*C_{EA},
\end{equation}
and the ratio of the natural structural frequency in vacuum to that in still fluid is given by
\begin{equation}\label{eq:fn_ratio}
	\frac{f_n}{f_{n,a}}=\sqrt{\frac{m^*+C_a}{m^*}}.
\end{equation}

\begin{figure}
\centerline{\includegraphics[width=0.75\textwidth]{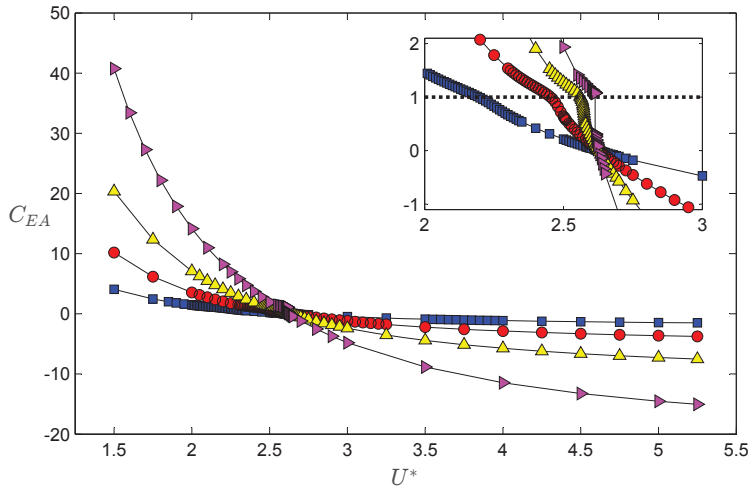}}
\caption{The variation of the effective added mass $C_{EA}$ with the reduced velocity $U^*$ for different mass ratios $m^*$ (symbol legend as in figure \ref{fig:forces}).} \label{fig:CEA} 
\end{figure}

In figure \ref{fig:CEA}, we present the variation of $C_{EA}$ as a function of $U^*$ for each $m^*$ investigated. $C_{EA}$ values were computed \textit{via} equation (\ref{eq:CGW}) and the known frequency ratio; note that  $f/f_{n,a}=f^*U_a^*=f^*U^*(f_n/f_{n,a})$. 
It can be seen that $C_{EA}$ decreases continuously  from positive to negative values with $U^*$; $C_{EA}$ decreases from as high as 40.8 to as low as $-13.3$ for $m^*=20$. At some point, $C_{EA}$ takes the theoretical $C_a$ value of unity, which is indicated by the dashed line in the inset of figure \ref{fig:CEA}. Interestingly, this occurs at approximately the reduced velocity of peak amplitude response, which is given by equation~(\ref{eq:Upeak}). 
Furthermore, for $m^*=20$ a gap   separating operating points with $C_{EA}$ values above and below unity appears at $U^*=2.614$, which is not present at lower $m^*$values. The discontinuous variation remained in place even though  an extremely fine step of $\Delta U^*=0.002$ was employed around this point. 
When the frequency of cylinder oscillation approaches the natural frequency of the structure in vacuum, equation~(\ref{eq:CGW}) yields $C_{EA}=0$, which occurs at the coincidence point $U^*=2.625$ for all mass ratios. 

The very wide variation of $C_{EA}$ values as a function of $U^*$  is improbable to represent some physical change, e.g.\ due to inertial effects, as we have already seen that flow physics remain fairly robust over the entire range of reduced velocities.  We  point  out  this simply because we would like to decipher the true effect due to the added mass, which seems impossible to do through the effective added mass concept. On the other hand, it is shown  further below with the aid of the new theory that a constant value of the added mass coefficient aligns very well the observations from the simulations.  
Nonetheless, the empirical values of $C_{EA}$ as defined above are valid.

\subsubsection{Super-harmonic fluid forcing at the coincidence point}\label{sec:superharmonic}

As already pointed out, the standard deviation of the unsteady in-line force tends to zero level as the frequency of cylinder vibration approaches the natural frequency of the structure in vacuum, which occurs at $U^*\approx2.625$ for all mass ratios. 
This can be predicted from the harmonic solution as follows. Equations (\ref{eq:Hsine}) and (\ref{eq:Hcosine}) given in appendix \ref{app:harmonic} can be combined so as to eliminate the phase angle $\phi_x$. In the case of zero structural damping $(\zeta=0)$,  the following relationship is obtained
\begin{equation}\label{eq:Ctotal}
	C_{x1} = 2\upi^3\frac{m^*A^*}{{U^*}^2}\left| 1 - \left(f^*U^*\right)^2 \right|.
\end{equation}
The above relationship shows that if the vibration frequency approaches the natural frequency of the structure in vacuum, which can be expressed in non-dimensional form as $f^*U^*\rightarrow1$, then the only feasible solution is that $C_{x1}\rightarrow0$. In the present simulations we have found that for all mass ratios $f$ approaches $f_n$ within less than 0.1\%, which was made possible  by using very fine steps of  $\Delta U^*=0.01$ or even less. 

\begin{figure}
\centerline{\includegraphics[width=0.99\textwidth]{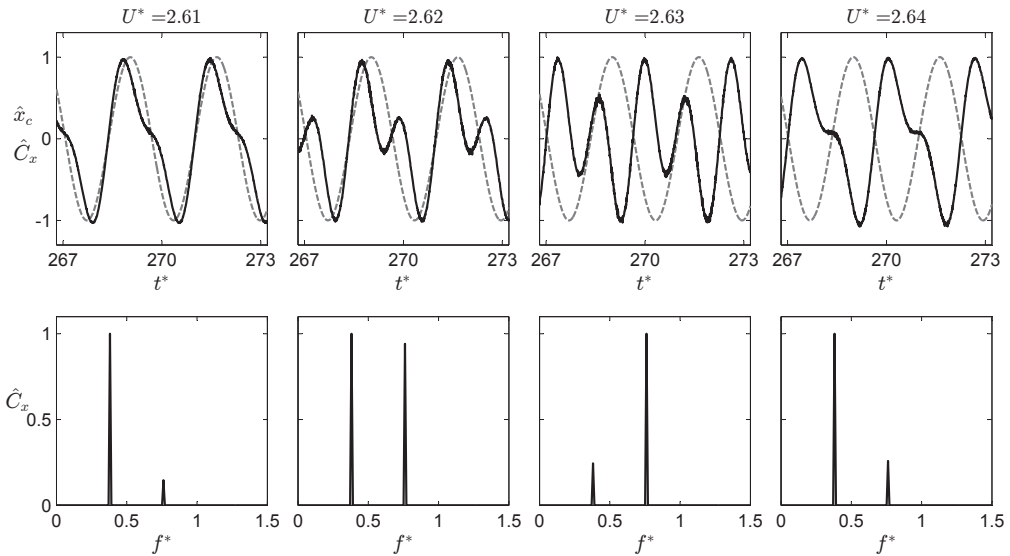}}
\caption{The top-row plots show time series of the cylinder displacement (dashed lines) and the unsteady streamwise force  (solid lines) for different reduced velocities around the coincidence point; $(Re,\,m^*)=(180,2)$. Here, the cylinder displacement $\hat{x}_c(t)$ and the in-line force $\hat{C}_x(t)$   have been normalized with their corresponding maximum amplitudes in order to reveal their relative waveforms.
The bottom-row plots show the spectra of the streamwise force for the corresponding  cases shown in the top row. } \label{fig:m2traces} 
\end{figure}

Figure \ref{fig:m2traces} shows time series of the cylinder displacement and the fluctuating part of the in-line force at different reduced velocities. The displacement and the force were  normalized with their corresponding maxima in order to reveal their relative waveforms because absolute values of the force are extremely small. Although the displacement remains remarkably harmonic, the force starts to  deviate  from pure harmonic as $U^*$ is varied with a fine step around the coincidence point. Spectra of the force show that a secondary peak appears at the first super-harmonic of the vibration frequency.  At $U^*=2.63$,  the main spectral peak  appears at the first super-harmonic of the vibration frequency, whereas a very small peak remains at the main frequency of vibration; the latter  corresponds to an extremely small magnitude of $C_{x1}$, which is however sufficient to sustain the vibration at the main harmonic of the vibration. 
The emergence of a dominant super harmonic accompanies the phase jump by $180^\circ$ of the main harmonic occurring as the coincidence point is crossed over, which can be also seen from the time traces in figure \ref{fig:m2traces}. The predominance of the first super harmonic was also observed in flow-induced transverse vibration of a rotating cylinder  at corresponding  coincidence points \citep{Bourguet2014}. It seems that the phase-jump mechanism through a super harmonic could be a generic feature.

\section{Development of new linear theory}\label{sec:theory}

In this section, we develop a theoretical framework  based on the triple decomposition of the total in-line force $F_x(t)$ in equation~(\ref{eq:Morison2}) and the reduced-order model of the wake force $F_{dw}(t)$ in equation~(\ref{eq:Fvharmonic}).  Initially, we  derive analytical expressions from which the model parameters $C_d$, $C_{dw}$ and $\phi_{dw}$ can be calculated using numerical data from the simulations. Then, we present the variations of the model parameters with reduced velocity and mass ratio to elucidate the fluid dynamics of vortex-induced in-line vibration. In addition, we combine the analytical expressions for drag components with the equation of cylinder motion in order to derive further expressions that allow us to interpret the flow physics at play. 

\subsection{Linearised force}
We begin with linearising the quasi-steady drag term in  equation~(\ref{eq:Morison2}) and substituting the harmonic approximations in equations (\ref{eq:Xharmonic})  and (\ref{eq:Fvharmonic}) to obtain the linearised force, which we can express  explicitly as a function of time $t$ as
\begin{eqnarray}
	F_x(t) = \frac{1}{2}\rho U_\infty^2D \left[ C_d +2 \left(\frac{\omega A}{U_\infty}\right) C_d \sin{(\omega t)}  +  C_{dw} \cos{(\omega t+\phi_{dw})} \right] \nonumber\\   +   \frac{1}{4}\upi\rho D^2C_a \omega^2A\cos{(\omega t)}, \label{eq:Morison3}
\end{eqnarray}
where the angular frequency $\omega=2\upi f$ has been introduced for brevity and $f_{dw}$ has been replaced by $f$ due to the synchronization of the wake and the cylinder vibration.

The first term on the right-hand side of equation (\ref{eq:Morison3}) represents the steady part of the in-line force whereas the remaining terms represent the unsteady part, which comprises the combination of cosine and sine functions at the frequency of cylinder oscillation. Equating the steady parts in equations (\ref{eq:Fharmonic})  and (\ref{eq:Morison3}) yields 
\begin{equation}\label{eq:meanCd}
			C_d  = \overline{C}_x, 
\end{equation}
i.e.\ $C_d$ is equal to the mean drag coefficient. Equating the cosine and sine terms in equations (\ref{eq:Fharmonic})  and (\ref{eq:Morison3}) yields the following relationships:
\begin{eqnarray}
	C_{dw}\sin{\phi_{dw}}  & = & C_{x1}\sin\phi_x + 4\upi f^*A^* C_d, \label{eq:Cdvsin} \\ 	
	C_{dw}\cos{\phi_{dw}} & = &  C_{x1}\cos\phi_x -  2\upi^3f^{*2}A^*C_a. \label{eq:Cdvcos}
\end{eqnarray}
 The set of equations (\ref{eq:meanCd}-\ref{eq:Cdvcos}) establishes relationships between fluid forcing components as functions of $A^*$ and $f^*$ alone, i.e.\ the structural parameters do not appear. Therefore, the same values of the fluid forcing components also  hold when the cylinder is forced to oscillate at the same operating points in the $A^*:f^*$ parameter space as the free vibration. We use the above set of relationships in order to determine $C_{dw}$ and $\phi_{dw}$ appearing in the new theoretical model from the cylinder response $(A^*,\,f^*)$ and  the fluid forcing $(C_{x1},\,\phi_{x})$ data obtained from the simulations. It should be remembered here that $C_a=1$ is the known ideal added-mass coefficient of unity.

\subsection{Magnitude and phase of wake drag}
\begin{figure}
\centerline{\includegraphics[width=0.7\textwidth]{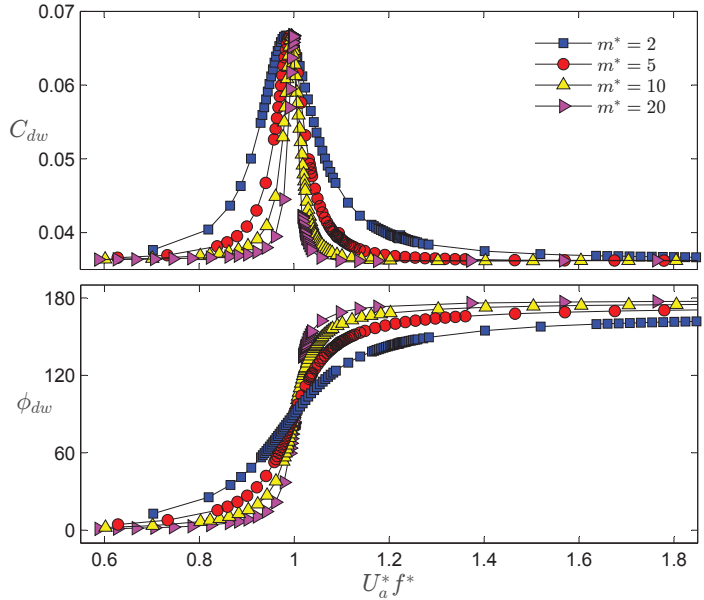}}
\caption{The variation of the wake drag magnitude $C_{dw}$ and phase $\phi_{dw}$ as functions of $U^*_af^*$ for different mass ratios $m^*$ at $Re=180$ (see the symbol legend for $m^*$ values).} \label{fig:vortex_drag} 
\end{figure}

Figure \ref{fig:vortex_drag}  shows the variation of $C_{dw}$ and  $\phi_{dw}$ as functions of the parameter $U^*_af^*$. We have employed  this parameter because peak amplitudes  occur at $U^*_af^*\approx1$. As can be seen in the top plot,  $C_{dw}$ also displays a peak  at the same point with  a maximum value of 0.067 for all $m^*$. The mutual amplification of the response amplitude and the wake drag are representative of resonance. Hence, we refer to the condition $U^*_af^*=1$ as the `resonance point'.  Away from resonance $C_{dw}$ tends asymptotically to 0.036, which corresponds to  the amplitude of the fluctuating drag coefficient for a non-vibrating cylinder at $Re=180$.   

\begin{figure}
\centerline{\includegraphics[width=0.74\textwidth]{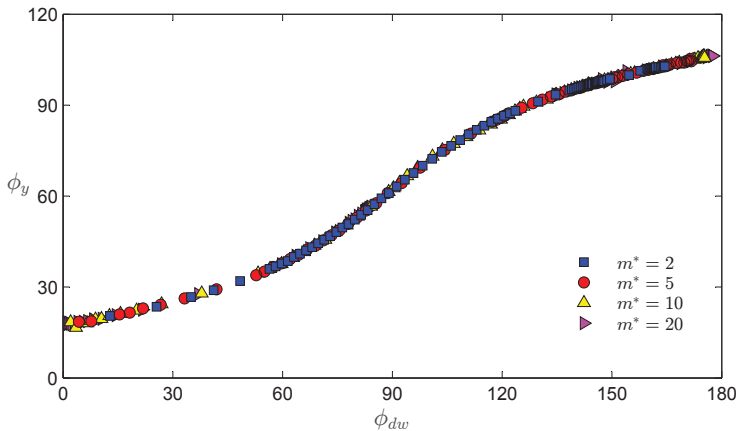}}
\caption{Relationship between phase angles $\phi_{dw}$ and $\phi_{y}$ for different mass ratios $m^*$ (see the symbol legend for $m^*$ values).} \label{fig:phaselink} 
\end{figure}

In the bottom plot in figure~\ref{fig:vortex_drag} can be seen that $\phi_{dw}$ follows an S-type increase as a function of $U^*_af^*$ from  approximately $0^\circ$ to $180^\circ$. At $U^*_af^*=1$, $\phi_{dw}$ passes through  $90^\circ$ for all $m^*$. 
Overall, the variation of  $\phi_{dw}$ is very similar to that of $\phi_y$ shown earlier; in fact there is a direct relationship between them, which is illustrated in figure \ref{fig:phaselink}. Since the variation of $\phi_y$ is invariably linked to the vortex dynamics, in particular to the timing of vortex shedding as discussed earlier, it follows that $\phi_{dw}$ appropriately captures related changes as the reduced velocity is varied. 

\subsection{Peak response at resonance point}
The new hydrodynamic model can be combined with the equation of cylinder motion in order to illustrate the interaction between the fluid and the structural dynamics. 
The steady-state solution can be obtained following a similar procedure as described in appendix \ref{app:harmonic}, which results in the following  two relationships: 
\begin{equation}
	C_{dw}\sin{\phi_{dw}}  =  4\upi f^*A^* \left(C_d + \frac{\upi^2m^*\zeta}{U^*}\right), \label{eq:Cdvsin2}\\
\end{equation}
\begin{equation}
	C_{dw}\cos{\phi_{dw}} =  2\upi^3\frac{m^*A^*}{U^{*2}}\left[1 - \left(\frac{f}{f_{n,a}}\right)^2  \right]. \label{eq:Cdvcos2}
\end{equation}
When the structural damping is zero $(\zeta=0)$, the solution of equation  (\ref{eq:Cdvsin2})   apparently does not depend on $m^*$. It should be noted however that the feasible operating points  in the $A^*:f^*$ space are limited to those that satisfy equation~(\ref{eq:Cdvsin2}), which correspond to the contour of zero energy transfer $(\sin\phi_x=0)$.  
In addition,  equation  (\ref{eq:Cdvcos2})  becomes indeterminate at the resonance point where $f=f_{n,a}$ and $\phi_{dw}=90^\circ$ simultaneously.   It should be noted that  there is no \textit{a priori} guarantee that the `resonance point' is attainable since the solution pursued here is open form. Nevertheless, as we can see from the simulations the resonance point is  attained and corresponds to the maximum amplitude, in which case we can write equation  (\ref{eq:Cdvsin2}) as
\begin{equation}\label{eq:peak}
	A^*_\mathrm{max} = \frac{C_{dw}}{4\upi f^*C_d}.
\end{equation}
This result shows that the steady-state solution at the resonance point does  not depend on $m^*$; rather the operating point is solely determined  by the fluid dynamics. 

When the structural damping is finite positive $(\zeta>0)$, equation~(\ref{eq:peak}) may still be used to estimate approximately the peak amplitude for  very low values of the mass and the damping such that $\upi^2m^*\zeta/U^* \ll C_d$, on the condition that the resonance point is  attained. The early experiments of \citet{Aguirre1977} support the above inference; he observed little variation of the peak amplitude in the second excitation region, i.e.\ the one associated with alternating vortex shedding,  which occurred at $U_a^*\approx 1.2/(2S)$ for $m^*=1.2$ and 4.3 (see figure 43 of his thesis). The mass--damping  was reported in terms of a stability parameter $k'_{s0}$   to be approximately 0.5, which was estimated from free-decay tests in still water, i.e.\ it includes contributions from hydrodynamic damping from the cylinder as well as its mountings. From his data, we estimate here that the corresponding $k_s$ value  in air was significantly lower than the one measured in water so that we can assume $\upi^2m^*\zeta/U^*<0.05$; this is small compared to the mean drag coefficient $C_d$. Thus, we argue that the present analysis is consistent with the experimental facts at  Reynolds numbers higher than considered in this study. 

At the resonance point  ($\phi_{dw}=90^\circ$), equations (\ref{eq:Cdvsin}) and (\ref{eq:Cdvcos})   reduce to 
\begin{eqnarray}
	C_{dw}  & = & 4\upi f^*A^* C_d, \label{eq:Cdvsin3}\\
 C_{x1} & = & 2\upi^3f^{*2}A^*C_a. \label{eq:Cdvcos3}
\end{eqnarray}
The above set of equations establishes relationships between the wake, mean, and total drag coefficients at resonance point, which  involve  variations in $A^*$ and  $f^*$. 
Considering that all three force coefficients, $C_{dw}$, $C_{d}$, and $C_{x1}$, may be `uniquely' specified in the parameter space $A^*:f^*$, this could suggest a single operating point inside this space at which a steady-state harmonic solution is feasible,  which implies that the cylinder peak response is drawn towards this operating point irrespectively of the mass ratio. Nonetheless, it should be cautioned that bimodal dynamics can also appear at particular regions of the parameter space $A^*:f^*$ \citep{Cagney2013pof,Cagney2013a,Gurian2019}. Assuming that a unique solution exists at the resonance point, equation~(\ref{eq:Cdvsin3}) illustrates that the component of the wake drag in-phase with the cylinder velocity counterbalances the quasi-steady drag, whereas  equation~(\ref{eq:Cdvcos3}) illustrates that the only contribution of the fluid force in-phase with the cylinder acceleration is the inviscid added-mass force. This is commensurate to the situation where an elastically mounted cylinder oscillates freely within a fluid medium with no net viscous forces; in such a situation, it would be anticipated that the frequency of cylinder oscillation is equal to the natural frequency of the structure in an inviscid fluid, i.e.\ $f=f_{n,a}$. This is  as if the fluid was phenomenologically interacting with the cylinder motion only through its ideal inertia.  It should be emphasized that in the case of a viscous fluid, the zero net viscous force results from the cancelling out of the wake drag and the quasi-steady drag, which is brought about by a gradual change in the phasing of the vortex shedding, thereby in $\phi_{dw}$, as the reduced velocity is varied. 

The above changes can be viewed from another perspective, which is more insightful. As $U^*$ increases, the phasing of vortex shedding gradually shifts to follow the changes in the frequency of oscillation with the reduced velocity. Since these variations occur in a continuous manner for the low-$Re$ cases considered here, a  point is reached where the timing of vortex shedding induces a wake drag exactly in-phase with the cylinder velocity, $\phi_{dw}=90^\circ$.  At this point, the net viscous force must become null (Eq.~\ref{eq:Cdvsin3}), whereas only the added mass contributes to the force in-phase with the cylinder acceleration (Eq.~\ref{eq:Cdvcos3}). It is important to note again that these changes are brought in entirely by the fluid dynamics.

\subsection{Infinitesimal net force at coincidence point}
At first glance, it seems counter-intuitive  that the cylinder  experiences almost no net force at the coincidence point. However, this can be explained by the cancelling out of the fluid-force components. It should be remembered that when the structural damping is zero $(\zeta=0)$ the linearised solution of the equation of cylinder motion (Eq.~\ref{eq:Ctotal}) shows that the component of the streamwise force at the main harmonic is null $(C_{x1}=0)$ at the coincidence point $(f^*U^*=1)$. In this case, equations (\ref{eq:Cdvsin})  and (\ref{eq:Cdvcos}) reduce to 
\begin{eqnarray}
	C_{dw}\sin{\phi_{dw}}  & = & 4\upi f^*A^* C_d, \label{eq:Cdvsin4} \\ 	
	-C_{dw}\cos{\phi_{dw}} & = & 2\upi^3f^{*2}A^*C_a. \label{eq:Cdvcos4}
\end{eqnarray}
From these equations, we can see that the wake drag in-phase with the velocity of the cylinder balances the quasi-steady drag while the wake drag in-phase with acceleration balances the inviscid inertia. This is physically possible as these force contributions originate from different flow structures. Furthermore, the cylinder must be oscillating for this case to be realized; the terms on the right-hand side of equations (\ref{eq:Cdvsin4}) and (\ref{eq:Cdvcos4}) are null for a non-oscillating cylinder. Although the above scenario is idealised as some damping will be present in a real situation, extending the analysis for a system with very low structural damping shows that the cylinder will experience a very low net in-line force at the coincidence point, if this point can still be  attained.  It should be noted that we approached very close to the coincidence point by employing a very fine step around this region, but  exact coincidence was unattainable. A similar observation was made by  \citet{Shiels2001} who showed that the net transverse force $C_y(t)$ exerted on a  cylinder undergoing free vibration in the transverse direction becomes null at limiting values of the structural parameters, $m=c=k=0$. Here, we show that  a null net streamwise force $C_x(t)$ on a cylinder undergoing free in-line vibration  is physically possible at the coincidence point for finite $m$ and $k$ values.

\subsection{$A^*_\mathrm{max}$ as a function of $Re$ in the laminar regime}

\begin{table} 
\begin{center}
\def~{\hphantom{0}}
\small\addtolength{\tabcolsep}{16pt}
\begin{tabular}{lcc}
$Re$ & $SC_{d}$   &  $(f^*/2)C_d$\\
100 & 0.217 & 0.216 \\
180 & 0.252 & 0.244 \\
\end{tabular}
\caption{\label{tab:SCd} Compilation of data for stationary and vibrating circular cylinders in the laminar regime. Data for the stationary cylinder are from  the study of \citet{Qu2013} and for the freely vibrating cylinder are  from the present simulations at peak amplitude.}
\end{center}
\end{table} 

\begin{figure}
\centerline{\includegraphics[width=0.82\textwidth]{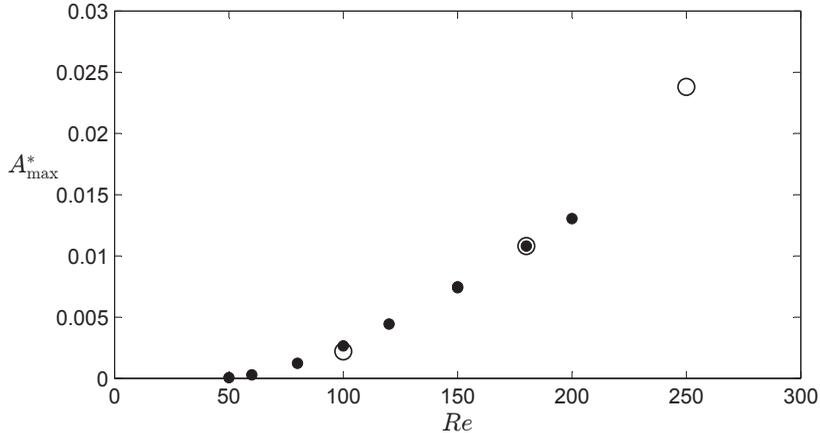}}
\caption{Comparison of $A^*_\mathrm{max}$ as a function of $Re$ obtained from the simulations (open symbols) and predicted from equation (\ref{eq:Apredict}) (full symbols).} \label{fig:Amax} 
\end{figure}

The variation of the peak amplitude as a function of the Reynolds  number can be estimated using the following assumptions. The mean drag coefficient $C_d$ and the inverse of the Strouhal number $1/S$ for stationary bluff bodies display similar variations as functions of $Re$  so that the product  $SC_d$ remains almost constant \citep{Alam2008}. For vibrating cylinders, there is also evidence that the corresponding product $(f^*/2)C_d$ remains constant \citep{Griffin1978}. \citet{Alam2008} showed that $SC_d\approx0.25$ for stationary bluff bodies of various cross-sectional shapes throughout the sub-critical regime. However, $SC_d$ appears to increase slightly with $Re$ in the laminar regime as seen in their data compilation (see their figure 5). We have confirmed this dependency on $Re$ in the laminar regime for the stationary  as well as the freely-vibrating circular cylinder as shown in table~\ref{tab:SCd}.  Nevertheless, the variation of $SC_d$ and  $(f^*/2)C_d$ with $Re$ is small, and in order to keep some generality in the result, we assume that $f^*C_d$ remains approximately constant. In this case, equation (\ref{eq:peak}) shows that $A^*_\mathrm{max}\propto C_{dw}$. In addition, we assume  $C_{dw}$ is magnified by some constant factor at resonance so that $A^*_\mathrm{max}\propto C_{dw0}$, where $C_{dw0}$ is the fluctuating drag coefficient for a stationary cylinder. Then, peak amplitudes at different $Re$ can be estimated from data at one particular Reynolds number $Re_0$, i.e.
\begin{equation}\label{eq:Apredict}
	A^*_\mathrm{max}(Re) = \left( \frac{A^*_\mathrm{max}}{C_{dw0}} \right)_{Re_0} C_{dw0}(Re).
\end{equation}
Figure \ref{fig:Amax} shows predictions of $A^*_\mathrm{max}$ using $C_{dw0}$ data at different $Re$ \citep[taken from][]{Qu2013} and the known response at $Re=180$.  It can be seen that $A^*_\mathrm{max}$ predictions  fit well with the amplitude from the simulations  at $Re=100$ while they also extrapolate to the result at $Re=250$. The enhancement of $A^*_\mathrm{max}$ with $Re$ is close to quadratic, which is quite marked compared to the nearly constant amplitude of purely transverse vortex-induced vibration in the laminar regime \citep[see][]{Govardhan2006}.

\section{Conclusions}
In this study we have developed a theoretical model for the fluid force acting on a circular cylinder vibrating in-line with a free stream. The streamwise fluid force comprises the sum of three components: an inviscid added-mass force opposing the inertia of the cylinder, a quasi-steady drag force opposing the velocity of the cylinder, and a wake drag force. 
The key element  here is the splitting of the viscous force into quasi-steady and wake components, which stem from the contribution of vorticity in fairly-separable flow regions, i.e.\ in the boundary/free shear layers and in the near wake, respectively. 

The theory enabled us to decipher three important  fluid-dynamical aspects. First, the phase angle of the vortex force with respect to the cylinder displacement $\phi_{dw}$ increases smoothly with $U^*$ due to a gradual shift in the timing of vortex shedding. This contrasts the variation of the phase angle  between the total force and the cylinder displacement, which must remain fixed at $\phi_x=0^\circ$ for $U^*f^*<1$ and  at $\phi_x=180^\circ$ for $U^*f^*>1$, with a sudden jump at the coincidence point $(U^*f^*=1)$; this constraint is imposed by the equation of cylinder motion when the structural  damping is null and has no bearing on the fluid dynamics.   Second, the added mass coefficient for inviscid flow is also applicable for viscous flow about a cylinder oscillating in-line with a free stream. Third, the wake drag is amplified in the excitation region of relatively high-amplitude response, which supports the classification of vortex-induced in-line vibration as a resonance phenomenon. 

The above findings could be confirmed because, unlike previous works dealing with vortex-induced vibration that primarily consider the fluid force in the direction of cylinder motion, here we calculated the phase angle between the  force transverse to the direction of motion and  the cylinder displacement, $\phi_y$. This allowed us to illustrate that the smooth variation of $\phi_y$ as a function of $U^*$, in contrast to $\phi_x$ that displays a sudden jump by $180^\circ$, is directly linked to a gradual shift in the  timing of vortex shedding. 

The theory developed in this investigation is linear, which means that all force components can be well approximated by single-harmonic functions of time, although the fluid dynamics is non linear by default. Departures from linearity can arise at much higher amplitudes and/or frequencies of cylinder vibration because of the quadratic relative velocity term. In addition, other non linearities may arise due to  competing requirements posed by the structural dynamics and the fluid dynamics. In fact, we have pinpointed in the present study that such non-linear effects become apparent at the `coincidence point' where the vibration frequency becomes equal to the natural frequency of the structure in vacuum. At this point, the equation of cylinder motion requires that the component of the in-line force at the main harmonic of the vibration becomes almost null, which results from balances between the quasi-steady drag and the wake drag in-phase with the velocity, and between the inviscid inertia and the wake drag in-phase with acceleration. In this case, the component at the first super-harmonic of the vibration frequency dominates the driving in-line force. 

An important result predicted from the theory is that the response does not depend on  $m^*$ at the point where $\phi_{dw}=90^\circ$. The simulations show that this occurs at the `resonance point' where the same peak amplitude is attained for all $m^*$ values, in accord with the theory. It should be noted however that the theory cannot predict that the maximum amplitude occurs when $\phi_{dw}=90^\circ$, or that this operating point does materialize, since the solution is given in open form. 
On the other hand, the flow physics suggest that at some point $\phi_{dw}=90^\circ$ due to the gradual shift in the timing of vortex shedding, which accompanies the continuous variation of the vibration frequency as the reduced velocity is varied. Nevertheless, we have also observed that discontinuities may arise by changing some parameter, e.g.\ $Re$ or $m^*$. Further study over wider ranges of these parameters than considered here may worth undertaking in the future.  

The simulations show that Reynolds number has a remarkable effect on the maximum amplitudes $A^*_\mathrm{max}$ attained over the entire reduced velocity range;  increasing $Re$ from 100 to 250 results in a  12-fold increase of $A^*_\mathrm{max}$. This stands in sharp contrast to the free vibration transversely to a free stream, in which case peak amplitudes remain fairly constant in the laminar regime. This may be attributable to variations of the added mass coefficient  in the latter configuration \citep{Konstantinidis2013prsa}. In view of this complexity, we consider  that free in-line vibration offers a more convenient test case to uncover the fluid dynamics.
The fluid dynamics could be elucidated because in-line response amplitudes remain very small for the low Reynolds  numbers investigated in the present study. As a consequence, the fluid excitation comes solely from the primary wake instability associated with alternating vortex shedding, which remains robust and similar as in the wake of a non-vibrating cylinder. It has been well established in the published literature  that other instabilities can be excited, even at small amplitudes, with increasing the Reynolds number,  such as the symmetrical mode of vortex shedding. There may also exist competition between different modes. Further instabilities due to gradual transition to turbulence will inevitably perplex the phenomenology of vortex-induced vibration. However, we maintain that the theory developed here remains valid and can be used to analyse the more complex phenomena  at higher Reynolds numbers, possibly with some adjustments for different fluid excitation mechanisms.  

\vspace{3pt}
{\bfseries Acknowledgements.} 
This research was supported by the European Union and the Hungarian State, co-financed by the European Regional Development Fund in the framework of the GINOP-2.3.4-15-2016-00004 project, aimed to promote the cooperation between the higher education and the industry.

\vspace{3pt}
{\bfseries Declaration of Interests.} The authors report no conflict of interest.

\appendix

\section{Steady-state harmonic solution}\label{app:harmonic}
 
Substitution of the harmonic approximations in Eqs. (\ref{eq:Xharmonic})  and (\ref{eq:Fharmonic}) into  the equation of cylinder motion (Eq.~\ref{eq:motion1}) and then balancing steady as well as unsteady sine and cosine terms on both sides of the resulting equation  yields the following three relationships:
\begin{eqnarray}
		kX_0 & = & F_{x0}, \\
		c\, \omega A & = & F_{x1}\sin{\phi_x},\\
		\left(-m\omega^2 + k \right) A & = & F_{x1}\cos{\phi_x},
\end{eqnarray}
where $\omega=2\upi f$ is the angular frequency of cylinder vibration. 

The relationships describing the steady-state periodic response may be rewritten in non-dimensional form as  
\begin{eqnarray}
		2\upi^3\,\frac{m^*}{U^{*2}}\,X_0^* & = & \overline{C}_x, \label{eq:Hmean}\\
		4\upi^3\frac{m^*\zeta}{U^*} f^*A^*& = & C_{x1}\sin{\phi_x},\label{eq:Hsine}\\
		2\upi^3\,\frac{m^*}{U^{*2}} \left[ 1 - \left(f^*U^*\right)^2\right] A^* & = & C_{x1}\cos{\phi_x}. \label{eq:Hcosine}
\end{eqnarray}

\section{Calculation of phase lag between forces and displacement}\label{app:phases}
The displacement of the cylinder and the fluid forces can be modelled to a very good degree of approximation (except near the coincidence point) as single-harmonic functions of time, i.e.\ their fluctuating parts can be written in non-dimensional form as
\begin{eqnarray}
	\hat{x}_c(t)=\cos{(2\upi ft')},\\
	\hat{C}_y(t)= \cos{(\upi ft'+\phi_y)},\\ \label{eq:CyA}
	\hat{C}_x(t)=  \cos{(2\upi ft'+\phi_x)}, \label{eq:CxA}
\end{eqnarray}
where hats above the symbols denote normalization with the corresponding maximum amplitudes, and $t'=t-t_0$ is the time with origin at a particular instant $t_0$ during the simulation where  the displacement oscillation is at peak after oscillations have stabilised to a steady state. It should be noted that in the present simulations the  frequency of alternating vortex shedding and frequency of the unsteady transverse force are both equal to half the frequency of cylinder oscillation (sub-harmonic synchronization).    
In the following, the procedure to calculate the phase angle $\phi_y$ of the transverse force with respect to displacement is shown in detail. Expansion of the cosine term in (\ref{eq:CyA}) yields $\hat{C}_y(t) = \cos\phi_y\cos{(\upi ft')} - \sin\phi_y\sin{(\upi ft')}$.
Multiplying $\hat{C}_y(t)$ by $\cos{(\upi ft')}$ and then taking the time average yields 
\begin{equation}\label{eq:Aone}
		\overline{\hat{C}_y(t)\cos{(\upi ft')}} = \cos\phi_y  \overline{\cos^2{(\upi ft')}} - \sin\phi_y  \overline{\sin{(\upi ft')}\cos{(\upi ft')}},
\end{equation}
where overlines denote time averaging. Similarly, multiplying $\hat{C}_y(t)$ by $\sin{(\upi ft')}$ and then taking the time average yields
\begin{equation}\label{eq:Atwo}
	\overline{\hat{C}_y(t)\sin{(\upi ft')}} = \cos\phi_y \ \overline{\cos{(\upi ft')}\sin{(\upi ft')}} - \sin\phi_y  \overline{\sin^2{(\upi ft')}}. 
\end{equation}
The time average of  $\overline{\sin{(\upi ft')}\cos{(\upi ft')}} $ over an even integer number of cycles is zero by default while $\overline{\cos^2{(\upi ft')}}=\overline{\sin^2{(\upi ft')}}$. Thus, substitution  of these values into (\ref{eq:Aone}) and (\ref{eq:Atwo}) and then combining them yields
\begin{equation}
	\phi_y = \tan^{-1}\left( \frac{-\overline{\hat{C}_y(t)\sin{(\upi ft')}}}
	{\overline{\hat{C}_y(t)\cos{(\upi ft')}}}  \right) .
\end{equation}
A similar procedure can be employed to compute the phase angle $\phi_x$ of the in-line force  by multiplying $\hat{C}_x(t)$ in Eq. (\ref{eq:CxA}) with $\cos{(2\upi ft')}$ and independently with $\sin{(2\upi ft')}$, averaging both equations and then combining them, which leads to the following result 
\begin{equation}
	\phi_x = \tan^{-1}\left( -\frac{\overline{\hat{C}_x(t)\sin{(2\upi ft')}}}
	{\overline{\hat{C}_x(t)\cos{(2\upi ft')}}}  \right).
\end{equation}
The accuracy of the above procedure is limited by the time step employed in the simulations, which is very small and thus leads to negligible errors in the calculation of the phase angles.

\bibliographystyle{jfm}
\bibliography{SVIV_paper_updated}

\end{document}